\documentclass[12pt,prb,aps]{revtex4-1}
\usepackage {amsmath}
\usepackage{amssymb}
\usepackage{stmaryrd}
\pdfoutput = 1 
\usepackage {graphicx,epstopdf}
\newcommand {\bxi}{\mbox{\boldmath$\xi$}}
\newcommand {\bta}{\mbox{\boldmath$\eta$}}
\allowdisplaybreaks

\begin{document}

\title{An Extended Variational Method for the Resistive Wall Mode in Toroidal Plasma Confinement Devices}
\author{R.~Fitzpatrick\,\footnote{rfitzp@utexas.edu}}
\affiliation{Institute for Fusion Studies,  Department of Physics,  University of Texas at Austin,  Austin TX 78712, USA}
\begin{abstract}
The external-kink stability of a toroidal plasma surrounded by a rigid resistive wall is investigated. 
The well-known analysis of Haney \& Freidberg is rigorously extended  to allow for a  wall that is sufficiently
thick that the thin-shell approximation does not necessarily hold. A generalized  Haney-Freidberg formula for the growth-rate of the resistive wall mode is obtained. Thick-wall effects do not
change the marginal stability point of the  mode, but introduce an interesting asymmetry between growing and decaying modes. Growing modes have
growth-rates that exceed those predicted by the original Haney-Freidberg formula. On the other hand, decaying modes have decay-rates that are less than those predicted by the original 
formula. 

The  well-known Hu-Betti formula for the rotational stabilization of the resistive wall mode is also generalized 
to take thick-wall effects into account. Increasing wall thickness  facilitates the
rotational stabilization of the  mode, because it decreases the critical toroidal electromagnetic  torque that the
wall must exert on the plasma. On the other hand, the real frequency of the mode at the marginal stability point increases with increasing wall thickness.
\end{abstract}
\maketitle

\section{Introduction}
According to standard ideal-magnetohydrodynamical (ideal-MHD) stability theory, a   fusion plasma confined on a set of toroidally nested magnetic flux-surfaces can be rendered completely stable to ideal external-kink
modes by means of a perfectly conducting, rigid wall that is located sufficiently close to the plasma boundary.\cite{ek1,ek2,ek3} Of course, a practical metal  
wall possesses 
a finite electrical conductivity, and 
can, therefore, only act as a perfect conductor on timescales that are much less than its characteristic L/R time. Given that the L/R time of any conceivable wall ($\approx 10^{-3}$\,s) is considerably smaller than
the desired confinement time of a fusion plasma ($\approx 1$ s),\cite{tok,fit} it is clear that the finite conductivity of the wall must be taken into account in the stability analysis. 
When the finite wall conductivity is taken into consideration, ideal external-kink modes that would be stabilized by the wall, were it perfectly conducting, are
found to grow on the L/R time of the wall.\cite{rwm1,rwm2}  Such comparatively slowly growing modes [compared to ideal external-kink modes, which grow on the extremely
short ($\approx 10^{-7}$\,s) Alfv\'{e}n time]\,\cite{bern,freid,freid1,goed} are known as {\em resistive wall modes}. In 1989, Haney \& Freidberg\,\cite{hf} derived a very general formula for the growth-rate of
a resistive wall mode  that makes use of the ``thin-shell approximation'', according to which the skin depth in the wall material is
assumed to be much larger than the wall thickness. The aim of this paper is to generalize the Haney-Freidberg formula to allow for thicker walls in
which the thin-shell approximation breaks down.\cite{thick0,thick01,chap,thick1,thick2,thick3,thick4} A secondary aim of the paper is to elucidate some subtleties in resistive wall
mode theory.  

\section{Ideal External-Kink Mode Stability}

\subsection{Introduction}
This section summarizes  some well-known background material.

\subsection{Scenrario}
\begin{figure}
\centerline{\includegraphics[width=0.6\textwidth]{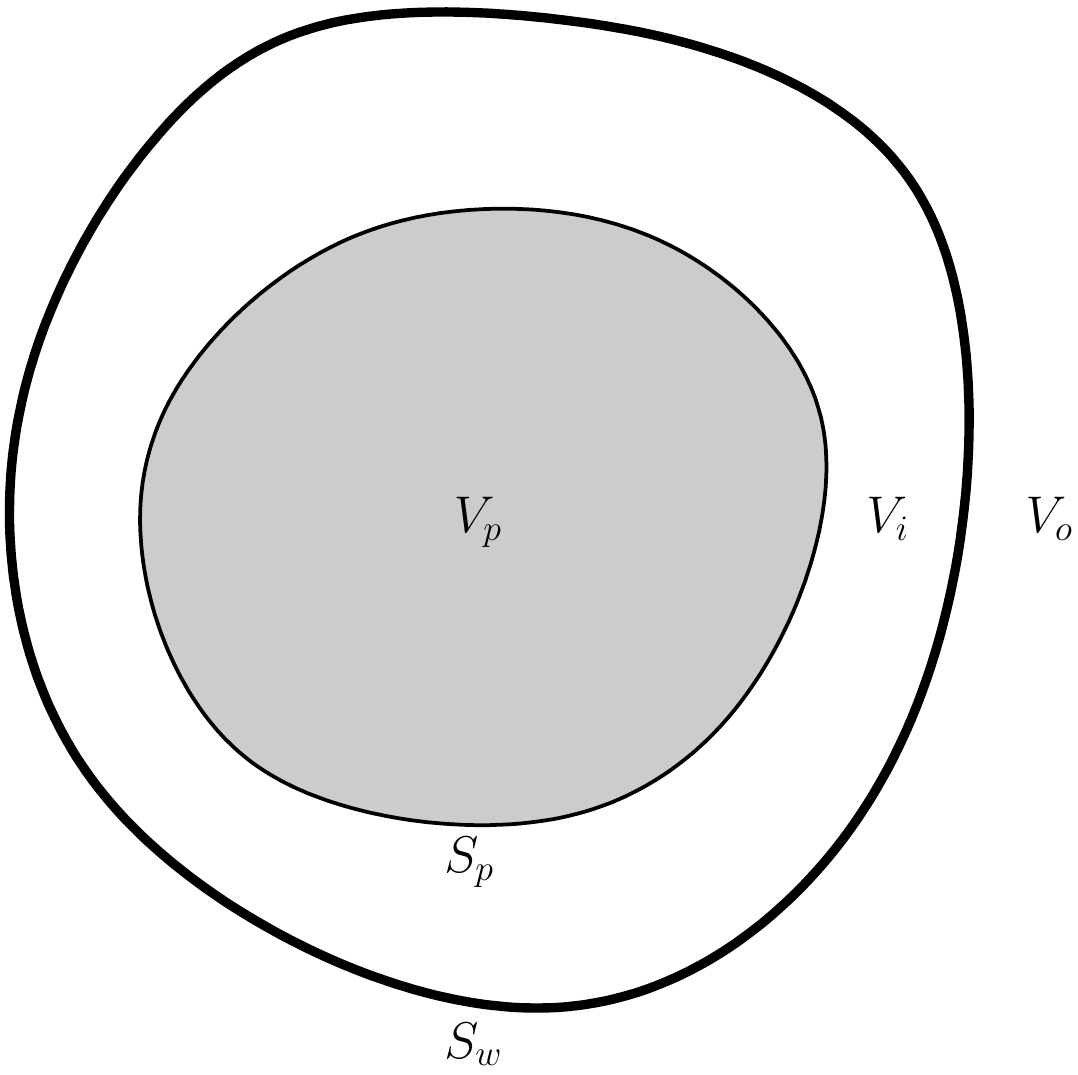}}
\caption{Schematic diagram showing the poloidal cross-section of a toroidally confined plasma. $V_p$ is the plasma volume, and 
$S_p$ is its bounding surface. $S_w$ is a physically thin wall that surrounds the plasma. $V_i$ is the vacuum region that lies between the plasma and the
wall, whereas $V_o$ is the vacuum region that lies outside the wall. }\label{fig1}
\end{figure}

Consider a fusion plasma that is confined on a set of toroidally-nested magnetic flux-surfaces. Let $V_p$ represent the toroidal volume occupied by the
plasma, and let $S_p$ be the volume's bounding surface. Suppose that the plasma is surrounded by a rigid, conducting wall whose uniform  thickness, $d$, 
is small compared to its effective minor radius, $\bar{b}$. Let the wall occupy the toroidal surface $S_w$. Let $V_i$ represent the vacuum region lying between the plasma boundary 
and the wall. Finally, let $V_o$ represent the vacuum region that lies outside the wall, and extends to infinity. See Fig.~\ref{fig1}. 

\subsection{Plasma Equilibrium}\label{sbeg}
Let $\rho({\bf r})$, $p({\bf r})$, ${\bf B}({\bf r})$, and ${\bf j}({\bf r})$ represent the equilibrium plasma mass density, scalar pressure, magnetic field,
and electric current density, respectively. It follows that $\mu_0\,{\bf j}=\nabla\times {\bf B}$, and $\nabla p = {\bf j}\times {\bf B}$. Let ${\bf n}$ be
a unit, outward directed normal vector to $S_p$. We have 
\begin{equation}\label{e1}
{\bf n}\cdot{\bf B}=0
\end{equation}
 on $S_p$, because $S_p$ must correspond to a magnetic flux-surface. We also expect
the rapid transport of particles and energy along magnetic field-lines to ensure that\,\cite{fit}
\begin{equation}
{\bf n}\times \nabla p=0
\end{equation}
 on $S_p$. Hence, we deduce from the equilibrium force balance equation 
that 
\begin{equation}\label{e3}
{\bf n}\cdot{\bf j}=0
\end{equation}
 on $S_p$. Finally, equilibrium force balance across $S_p$ yields
\begin{equation}
\left\llbracket p + \frac{B^2}{2\,\mu_0}\right\rrbracket\equiv \frac{\hat{B}^{2}}{2\,\mu_0}-p-\frac{B^2}{2\,\mu_0} =0
\end{equation}
on $S_p$, which implies that
\begin{equation}\label{e4}
\left\llbracket \nabla\cdot\left(p + \frac{B^2}{2\,\mu_0}\right)\right\rrbracket= \left\llbracket{\bf n}\cdot \nabla\left(p + \frac{B^2}{2\,\mu_0}\right)\right\rrbracket{\bf n}
\end{equation}
on $S_p$. Here, $\hat{\bf B}({\bf r})$ is the equilibrium magnetic field in the vacuum region. 

\subsection{Plasma Perturbation}
Assuming an $\exp(\gamma\,t)$ time dependence of all perturbed quantities, and neglecting equilibrium plasma flows, the perturbed, linearized plasma equation of motion takes
the form\,\cite{bern,freid,freid1,goed}
\begin{equation}\label{e6}
\gamma^2\,\rho\,\bxi = {\bf F}(\bxi),
\end{equation}
where
\begin{equation}
{\bf F}(\bxi) = \nabla({\mit\Gamma}\,p\,\nabla\cdot\bxi) -\mu_0^{-1}\,{\bf B}\times(\nabla\times {\bf Q}) +\nabla(\bxi\cdot\nabla p) + {\bf j}\times{\bf Q},
\end{equation}
and 
\begin{equation}
{\bf Q} =\nabla\times(\bxi\times {\bf B}).
\end{equation}
Here, $\bxi({\bf r})$ is the plasma displacement, ${\mit\Gamma}=5/3$ the ratio of specific heats, and ${\bf Q}({\bf r})$ the divergence-free perturbed magnetic field in the plasma. 
Moreover, ${\bf F}(\bxi)$ is known as the {\em force operator}. 
The divergence-free perturbed magnetic field in the vacuum region is written $\nabla\times {\bf A}$, where
\begin{equation}\label{e9}
\nabla\times (\nabla\times {\bf A}) = {\bf 0}.
\end{equation}

Now, $\bxi({\bf r})$ must be square integrable at the magnetic axis, otherwise the potential energy, $\delta W$ [see Eq.~(\ref{e19a})], of the perturbation would be infinite. 
Furthermore,\cite{freid,freid1,goed}
\begin{align}
{\bf n}\times {\bf A} &= -({\bf n}\cdot\bxi)\,\hat{\bf B},\label{e11}\\[0.5ex]
-{\mit\Gamma}\,p\,\nabla\cdot{\bxi} +\bxi\cdot\nabla\left(\frac{B^2}{2\,\mu_0}\right) + \mu_0^{-1}\,{\bf B}\cdot{\bf Q} 
&=\bxi\cdot\nabla\left(\frac{\hat{B}^2}{2\,\mu_0}\right) +\mu_0^{-1}\,\hat{\bf B}\cdot\nabla\times {\bf A}\label{e12}
\end{align}
on $S_p$. 
Equation~(\ref{e11}) ensures that the perturbed plasma boundary remains a magnetic flux-surface, whereas Eq.~(\ref{e12}) is an
expression of perturbed pressure balance across the boundary. Finally, if the wall is perfectly conducting then
\begin{equation}\label{e13}
{\bf n}_w\times {\bf A} = {\bf 0}
\end{equation}
on $S_w$, where ${\bf n}_w$ is a unit, outward directed normal vector to $S_w$. On the other hand, if the wall is absent then
\begin{equation}\label{e14}
{\bf A} = {\bf 0}
\end{equation}
at infinity. Equation~(\ref{e13}) ensures that the perturbed magnetic field cannot penetrate the perfectly conducting wall, whereas
Eq.~(\ref{e14}) ensures that the potential energy of the perturbation remains finite. 
The boundary conditions (\ref{e11}), (\ref{e13}), and (\ref{e14}), as well as the constraint that 
$\bxi({\bf r})$ be square integrable at the magnetic axis,  and the constraint that $\nabla\times{\bf A}$ be square integrable at infinity,  are conventionally termed {\em essential boundary conditions}, whereas
the boundary condition (\ref{e13}) is termed a {\em natural boundary condition}.\cite{freid,freid1,goed} An essential
boundary condition is one that must be satisfied by all prospective solution pairs, [$\bxi({\bf r})$, ${\bf A}({\bf r})$]. On the other hand, a natural boundary condition is one
that must be satisfied by physical solution pairs, but, as we shall see,  can be violated by trial solution pairs. 

\subsection{Self-Adjoint Property of Force Operator}\label{self}
Let [$\bxi({\bf r})$, ${\bf A}({\bf r})$] and [$\bta({\bf r})$, ${\bf C}({\bf r})$] be two independent prospective solution pairs that satisfy both the essential and the
natural boundary conditions. It is easily demonstrated that\,\cite{goed}
\begin{align}\label{e17}
 \int_{V_p}\bta\cdot {\bf F}(\bxi)\,dV_p&=-  \int_{V_p}\left[{\mit\Gamma}\,p\,(\nabla\cdot\bxi)\,(\nabla\cdot\bta)+\mu_0^{-1}\,{\bf Q}\cdot{\bf R}+\frac{1}{2}\,\nabla p\cdot[(\nabla\cdot\bta)\,\bxi
+(\nabla\cdot\bxi)\,\bta]\right.\nonumber\\[0.5ex]
&\phantom{=}\left.+\frac{1}{2}\,{\bf j}\cdot(\bxi\times {\bf R}+\bta\times {\bf Q})\right]dV_p \nonumber\\[0.5ex]
 &\phantom{=}+ \int_{S_p} ({\bf n}\cdot\bta)\left({\mit\Gamma}\,p\,\nabla\cdot\bxi + \bxi\cdot\nabla p - \mu_0^{-1}\,{\bf B}\cdot{\bf Q}\right)dS_p,
\end{align}
where ${\bf R}=\nabla\times(\bta\times {\bf B})$, and use has been made of Eqs.~(\ref{e1})--(\ref{e3}). 
The surface integral can be expressed as 
\begin{align}
 &\int_{S_p} ({\bf n}\cdot\bta)\left({\mit\Gamma}\,p\,\nabla\cdot\bxi + \bxi\cdot\nabla p - \mu_0^{-1}\,{\bf B}\cdot{\bf Q}\right)d{S_p}\nonumber\\[0.5ex]
&= -\int_{S_p}  ({\bf n}\cdot\bta)\,({\bf n}\cdot\bxi)\,{\bf n}\cdot \left\llbracket\nabla\left(p+ \frac{{B}^{2}}{2\,\mu_0}\right)\right\rrbracket d{S_p} +\mu_0^{-1}\int_{S_p}
{\bf n}\cdot {\bf C}\times (\nabla\times {\bf A})\,d{S_p},
\end{align}
where use has been made of Eqs.~(\ref{e4}), (\ref{e11}), and (\ref{e12}). 
However, the boundary conditions (\ref{e13}) and (\ref{e14}) allow us to write
\begin{equation}
\int_{S_p}
{\bf n}\cdot {\bf C}\times (\nabla\times {\bf A})\,d{S_p} = -\int_{V}\nabla\cdot[{\bf C}\times(\nabla\times {\bf A})]\,dV=-\int_V
(\nabla\times {\bf C})\cdot(\nabla\times {\bf A})\,dV,
\end{equation}
where use has been made of Eq.~(\ref{e9}). Here, $V$ denotes the appropriate vacuum region: i.e., $V_i$ for the case of a perfectly
conducting wall, and $V_{io}=V_i+V_o$ for the case of no wall. The previous three equations yield 
\begin{align}\label{e18}
 \int_{V_p} \bta\cdot{\bf F}(\bxi)\,dV_p&=-  \int_{V_p}\left[{\mit\Gamma}\,p\,(\nabla\cdot\bxi)\,(\nabla\cdot\bta)+\mu_0^{-1}\,{\bf Q}\cdot{\bf R}+\frac{1}{2}\,\nabla p\cdot[(\nabla\cdot\bta)\,\bxi
+(\nabla\cdot\bxi)\,\bta]\right.\nonumber\\[0.5ex]
&\phantom{=}\left.+\frac{1}{2}\,{\bf j}\cdot(\bxi\times {\bf R}+\bta\times {\bf Q})\right]d{V_p}\nonumber\\[0.5ex]
&\phantom{=} -\int_{S_p}  ({\bf n}\cdot\bta)\,({\bf n}\cdot\bxi)\,{\bf n}\cdot \left\llbracket\nabla\left(p+ \frac{B^{2}}{2\,\mu_0}\right)\right\rrbracket d{S_p}
\nonumber\\[0.5ex]
&\phantom{=} -\int_{V}\mu_0^{-1}\,(\nabla\times {\bf C})\cdot(\nabla\times {\bf A})\,dV.
\end{align}
The well-known self-adjoint property of the force operator,\cite{bern,freid,freid1,goed}
\begin{equation}\label{e19}
\int_{V_p} \bta\cdot{\bf F}(\bxi)\,d{V_p}= \int_{V_p} \bxi\cdot {\bf F}(\bta)\,d{V_p},
\end{equation}
immediately follows from the invariance of Eq.~(\ref{e18}) under the transformation $\bxi$, ${\bf A}$, ${\bf Q}\leftrightarrow \bta$, ${\bf C}$, ${\bf R}$, respectively. 

\subsection{Ideal-MHD Energy Principle}
The potential energy of the perturbation characterized by the solution pair  [$\bxi({\bf r})$, ${\bf A}({\bf r})$] takes the form\,\cite{bern,freid,freid1,goed}
\begin{equation}\label{e19a}
\delta W(\bxi,\bxi)= -\frac{1}{2}\int_{V_p}\bxi\cdot{\bf F}(\bxi)\,dV_p.
\end{equation}
It follows from Eq.~(\ref{e18}) that
\begin{equation}\label{e21}
\delta W = \delta W_p + \delta W_s + \delta W_v,
\end{equation}
where
\begin{align}
\delta W_p(\bxi,\bxi) &= \frac{1}{2}\int_{V_p}\left[{\mit\Gamma}\,p\,(\nabla\cdot\bxi)\,(\nabla\cdot\bxi)+\mu_0^{-1}\,{\bf Q}\cdot{\bf Q} + (\nabla\cdot\bxi)\,(\bxi\cdot\nabla p)
+ {\bf j}\cdot\bxi\times {\bf Q}\right]d{V_p},\\[0.5ex]
\delta W_s(\bxi,\bxi) &= \frac{1}{2}\int_{S_p} ({\bf n}\cdot\bxi)\,({\bf n}\cdot\bxi)\,{\bf n}\cdot\left\llbracket\nabla\left(p
+\frac{B^{2}}{2\,\mu_0}\right)\right\rrbracket d{S_p},\\[0.5ex]
\delta W_v({\bf A},{\bf A}) &= \frac{1}{2\,\mu_0}\int_{V}(\nabla\times {\bf A})\cdot(\nabla\times {\bf A})\,dV.
\end{align}
Clearly, $\delta W_p$, $\delta W_s$, and $\delta W_v$ respectively represent the contributions from the bulk plasma, from equilibrium surface currents flowing on the plasma boundary, and from
the vacuum, to the overall potential energy of the perturbation. 

The {\em ideal-MHD energy principle}\,\cite{bern,freid,freid1,goed} states that if any solution pair that satisfies the boundary conditions
makes $\delta W<0$ then the plasma is ideally unstable. In other words, at least one eigenmode with $\gamma^2>0$ exists, where $\gamma$ is of
order the inverse Alfv\'{e}n time. On the other hand, if no valid solution pair can be found such that $\delta W<0$ then the
plasma is ideally stable. In other words, all eigenmodes are characterized by $\gamma^2<0$. 

Clearly, in order to utilize the ideal-MHD principle, we need to minimize $\delta W(\bxi,\bxi)$ with respect to $\bxi$, and then determine whether the
minimum value is positive or negative. In the former case, the plasma is ideally stable. In the latter case, it is ideally unstable. 
We can write
\begin{equation}
\delta[\delta W(\bxi,\bxi)]= \delta W(\delta\bxi,\bxi)+ \delta W(\bxi,\delta\bxi) = 2\,\delta W(\delta\bxi,\bxi),
\end{equation}
where use has been made of the self-adjoint property of the force operator, (\ref{e19}). It is easily
demonstrated that 
\begin{align}\label{e25}
2\,\delta W_p(\delta\bxi,\bxi)&=-\int_{V_p}\delta\bxi\cdot{\bf F}(\bxi)\,dV_p\nonumber\\[0.5ex]
&\phantom{=}+ \int_{S_p}
({\bf n}\cdot\delta\bxi)\left({\mit\Gamma}\,p\,\nabla\cdot\bxi +\bxi\cdot\nabla p -\mu_0^{-1}\,{\bf B}\cdot{\bf Q}\right)dS_p,\\[0.5ex]
2\,\delta W_s(\delta\bxi,\bxi)&= \int_{S_p}({\bf n}\cdot\delta\bxi)\,\bxi\cdot\nabla\left(\frac{\hat{B}^{2}}{2\,\mu_0} - \frac{B^{2}}{2\,\mu_0}-p\right)d{S_p},\label{e26}\\[0.5ex]
2\,\mu_0\,\delta W_v(\delta{\bf A},{\bf A})
&= \int_{S_p}({\bf n}\cdot\delta\bxi) \,\hat{\bf B}\cdot(\nabla\times{\bf A})\,d{S_p}+ \int_{V}\delta{\bf A}\cdot\nabla\times(\nabla\times{\bf A})\,dV.
\end{align}
Here, use has been made of the boundary conditions (\ref{e11}), (\ref{e13}),  and (\ref{e14}), which are assumed to apply to $\delta\bxi$ and
$\delta{\bf A}$. Thus, setting $\delta[\delta W(\bxi,\bxi)]=0$, we get 
\begin{align}
0=&-\int_{V_p}\delta\bxi\cdot{\bf F}({\bxi})\,d{V_p}\nonumber\\[0.5ex]
&-\int_{S_p}({\bf n}\cdot\delta\bxi)\left[-{\mit\Gamma}\,p\,\nabla\cdot\bxi+\bxi\cdot\nabla\left(\frac{B^{2}}{2\,\mu_0}\right)
+\mu_0^{-1}\,{\bf B}\cdot{\bf Q} \right.\nonumber\\[0.5ex]&\phantom{=}\left.-\bxi\cdot\nabla\left(\frac{\hat{B}^{2}}{2\,\mu_0}\right)-\mu_0^{-1}\,\hat{\bf B}\cdot(\nabla\times{\bf A})\right]d{S_p}
\nonumber\\[0.5ex]
&+ \int_{V}\mu_0^{-1}\,\delta{\bf A}\cdot\nabla\times (\nabla\times{\bf A})\,dV.
\end{align}
Because the previous equation must hold for arbitrary $\delta\bxi$ and $\delta{\bf A}$, we deduce that the trial solution pair that
minimizes $\delta W$ satisfies the force-balance equation, 
\begin{equation}\label{e30}
{\bf F}(\bxi)= {\bf 0}, 
\end{equation}
in $V_p$, satisfies Eq.~(\ref{e9}) in $V$, and also ought to satisfy the pressure balance matching condition, (\ref{e12}), at the plasma boundary. 

\subsection{Perfect-Wall and No-Wall Stability}\label{send}
Suppose that the wall is perfectly conducting. In this case, the previous analysis suggests that the minimum value of the perturbed
potential energy can be written 
\begin{equation}\label{e30s}
\delta W_{pw} = \delta W_p + \delta W_s+ \delta W_v^{(b)},
\end{equation}
where $\delta W_p$ and $\delta W_s$ are calculated from a particular solution of Eq.~(\ref{e30}) that is well-behaved at the magnetic axis, and
\begin{equation}
\delta W_v^{(b)} = \frac{1}{2\,\mu_0}\int_{V_i} (\nabla\times{\bf A}_{pw})^2\,d{V_i}.
\end{equation}
Here, the superscript $(b)$ implies that the perfectly conducting wall is present at effective  minor radius $\bar{b}$. Moreover, ${\bf A}_{pw}({\bf r})$ is a solution of Eq.~(\ref{e9}) that satisfies the boundary conditions 
\begin{equation}\label{e32} 
{\bf n}\times {\bf A}_{pw} = -({\bf n}\cdot\bxi)\,\hat{\bf B}
\end{equation}
on $S_p$ [see Eq.~(\ref{e11})], and 
\begin{equation}\label{e33}
{\bf n}_w\times {\bf A}_{pw} = {\bf 0}
\end{equation}
on $S_w$. [See Eq.~(\ref{e13}).] We can be sure that if $\delta W_{pw}>0$ then the ideal external-kink mode in question is stabilized by the wall. 

Suppose that there is no wall. In this case, the previous analysis suggests that the minimum value of the perturbed
potential energy can be written 
\begin{equation}\label{e34}
\delta W_{nw} = \delta W_p + \delta W_s+ \delta W_v^{(\infty)},
\end{equation}
where $\delta W_p$ and $\delta W_s$ are calculated from the same solution of Eq.~(\ref{e30})  as that used to calculate $\delta W_{pw}$, and
\begin{equation}
\delta W_v^{(\infty)} = \frac{1}{2\,\mu_0}\int_{V_{io}} (\nabla\times{\bf A}_{nw})^2\,d{V_{io}}.
\end{equation}
Here, the superscript $(\infty)$ implies that the perfectly conducting wall is absent, which is equivalent to it being
placed at infinity. 
Moreover, ${\bf A}_{nw}({\bf r})$ is a solution of Eq.~(\ref{e9}) that satisfies the boundary conditions 
\begin{equation}\label{e36}
{\bf n}\times {\bf A}_{nw} = -({\bf n}\cdot\bxi)\,\hat{\bf B}
\end{equation}
on $S_p$ [see Eq.~(\ref{e11})], and 
\begin{equation}\label{exx}
{\bf A}_{nw} = {\bf 0}
\end{equation}
at infinity. [See Eq.~(\ref{e14}).] We can be sure that if $\delta W_{nw}<0$ then the ideal external-kink mode in question is unstable in the absence of the wall. 
Of course, the situation in which a kink mode is unstable in the absence of the wall, and stable in the presence of perfectly
conducting wall, is exactly that which pertains to the resistive wall mode. 

\subsection{Discussion}\label{s1d}
The reason that we have reproduced the very standard theory outlined in Sects.~\ref{sbeg}--\ref{send} is to make an important observation. 
Namely, the solution pairs, [$\bxi({\bf r})$, ${\bf A}_{pw}({\bf r})$] and  [$\bxi({\bf r})$, ${\bf A}_{nw}({\bf r})$], conventionally used to calculate
$\delta W_{pw}$ and $\delta W_{nw}$, respectively, do not satisfy the pressure balance matching  condition, (\ref{e12}), at the plasma boundary. One way
of seeing this is to note that if ${\bf F}(\bxi)={\bf 0}$ in $V_p$ then, according to Eq.~(\ref{e19a}),  $\delta W=0$. However, if
we assume that the pressure balance matching  condition is not satisfied then Eq.~(\ref{e21}) generalizes to
\begin{equation}\label{e38}
\delta W = \delta W_p + \delta W_s + \delta W_v + \delta W_c,
\end{equation}
where
\begin{align}\label{e39}
\delta W_c(\bxi,{\bf A})&= \frac{1}{2}\int_{S_p}({\bf n}\cdot\bxi)\left[-{\mit\Gamma}\,p\,\nabla\cdot\bxi+\bxi\cdot\nabla\left(\frac{B^{2}}{2\,\mu_0}\right)
+\mu_0^{-1}\,{\bf B}\cdot{\bf Q} \right.\nonumber\\[0.5ex]&\phantom{=}\left.-\bxi\cdot\nabla\left(\frac{\hat{B}^{2}}{2\,\mu_0}\right)-\mu_0^{-1}\,\hat{\bf B}\cdot(\nabla\times{\bf A})\right]d{S_p}.
\end{align}
Here, the surface energy $\delta W_c$ is directly related to the failure to satisfy the pressure balance matching condition at the plasma boundary. 
Thus, it is clear that
\begin{align}
\delta W_{pw} &= -\delta W_c(\bxi,{\bf A}_{pw}),\\[0.5ex]
\delta W_{nw}&= -\delta W_c(\bxi,{\bf A}_{nw}).
\end{align}
In other words, the only reason that $\delta W_{pw}$ and $\delta W_{nw}$ can take non-zero values at all is because the pressure balance matching condition
is not satisfied. See Sect.~\ref{sno}. 

The fact that the the solution pairs, [$\bxi({\bf r})$, ${\bf A}_{pw}({\bf r})$] and  [$\bxi({\bf r})$, ${\bf A}_{nw}({\bf r})$], do not satisfy the
pressure balance matching condition, (\ref{e12}), at the plasma boundary is not problematic within the context of ideal-MHD theory. The ideal-MHD energy principle
guarantees that if $\delta W_{nw}<0$ then we can find a solution of Eq.~(\ref{e6}) inside the plasma, and Eq.~(\ref{e9}) outside the
plasma, which satisfies all of the boundary conditions in the absence of a wall, and is such that $\gamma^2>0$. Likewise, if $\delta W_{pw}>0$ then we can find a solution of Eq.~(\ref{e6}) inside the plasma, and Eq.~(\ref{e9}) outside the
plasma, which satisfies all of the boundary conditions in the presence of a perfectly conducting wall, and is such that $\gamma^2<0$. 
In both cases, the undetermined Alfv\'{e}nic growth-rate of the instability, $\gamma$, provides the additional degree of freedom that allow
the pressure balance matching condition, (\ref{e12}), to be satisfied at the plasma boundary. 
However, the analysis of Haney-Freidberg constructs the resistive wall mode solution from a linear combination of  [$\bxi({\bf r})$, ${\bf A}_{pw}({\bf r})$] and  [$\bxi({\bf r})$, ${\bf A}_{nw}({\bf r})$]. Moreover, such a solution must satisfy all of the boundary conditions, including Eq.~(\ref{e12}). Thus, it is important to
demonstrate that this is actually possible. 

\section{Resistive Wall Mode Physics}

\subsection{Resistive Wall Physics}
Let ${\bf A}_i({\bf r})$ be the vector potential in $V_i$. Let ${\bf A}_o({\bf r})$ be the vector potential in $V_o$. Finally,
let ${\bf A}_w({\bf r})$ be the vector potential inside the wall. Choosing the Coulomb gauge within the wall,\cite{jackson} the electric
field in the wall takes the form
\begin{equation}
{\bf E}_w=-\gamma\,{\bf A}_w, 
\end{equation}
whereas the magnetic field is given by
\begin{equation}
{\bf B}_w = \nabla\times {\bf A}_w.
\end{equation}
Ohm's law inside the wall yields
\begin{equation}
{\bf j}_w= \sigma_w\,{\bf E}_w,
\end{equation}
where $\sigma_w$ is the uniform electrical conductivity of the wall material, and ${\bf j}_w$ is the density of the electrical current flowing in the wall. Finally,
\begin{equation}
\mu_0\,{\bf j}_w=\nabla\times {\bf B}_w.
\end{equation}
The previous four equations can be combined to give
\begin{equation}\label{e46}
\nabla\times (\nabla\times{\bf A}_w) = -\mu_0\,\sigma_w\,\gamma\,{\bf A}_w.
\end{equation}

Let $d$ and $\bar{b}$ be the uniform thickness and effective minor radius of the wall, respectively. Now, we are assuming that the wall is
physically thin: i.e., $d\ll \bar{b}$. 
Following Haney \& Freidberg,\cite{hf}  the position vector of a point lying within the wall is written
\begin{equation}
{\bf r} = {\bf r}_i + u\,d\,{\bf n}_w.
\end{equation}
where ${\bf r}_i$ is the position vector of a point on the inner surface of the  wall. The normalized length $u$ represents perpendicular distance measured outward
from the inner surface of the wall. Thus, $u = 0$ and $u = 1$ correspond to the inner and outer surface of the wall, respectively. 
We deduce  that the gradient operator within the wall can be written
\begin{equation}
\nabla = \frac{{\bf n}_w}{d}\,\frac{\partial}{\partial u} + \nabla_{S_w},\label{e48}
\end{equation}
where $\nabla_{S_w}\sim\bar{b}^{-1}$ only involves derivatives tangent to the surface of the wall.

To lowest order in $d/\bar{b}$, Eqs.~(\ref{e46}) and (\ref{e48}) yield
\begin{equation}\label{e49}
{\bf n}_w\times\left({\bf n}_w\times\frac{\partial^2\,{\bf A}_w}{\partial u^2}\right)=\lambda^2\,{\bf A}_w,
\end{equation}
where
\begin{equation}
\lambda = \sqrt{\mu_0\,\sigma_w\,d^{\,2}\,\gamma}.
\end{equation}
Let 
\begin{equation}\label{e51}
\delta_\lambda= \frac{d}{\lambda\,\bar{b}}.
\end{equation}
The neglect of tangential derivates with respect to perpendicular derivatives in Eq.~(\ref{e49}) is valid as long as
\begin{equation}
|\delta_\lambda|\ll 1,
\end{equation}
which we shall assume to be the case.  Note that
\begin{equation}
|\lambda| = \frac{d}{d_{skin}},
\end{equation}
where 
\begin{equation}
d_{skin} = \frac{1}{\sqrt{\mu_0\,\sigma_w\,|\gamma|}}
\end{equation}
is the skin depth in the wall material.\cite{jackson} The so-called {\em thin-shell}\/ limit correspond to the situation in which the skin depth is much
greater than the wall thickness: i.e., $|\lambda|\ll 1$. On the other hand, the {\em thick-shell}\/ limit corresponds to the situation in
which the skin depth is greater than or comparable to the wall thickness: i.e., $|\lambda|\gtrsim 1$. Note that it is possible
for the wall to be physically thin (i.e., $d/\bar{b}\ll 1$) but still be in the thick-shell limit. 

Equation~(\ref{e49}) yields
\begin{equation}
\frac{\partial^2}{\partial u^2}\,{\bf n}_w\times {\bf A}_w = \lambda^2\,{\bf n}_w\times {\bf A}_w.
\end{equation}
The solution is
\begin{equation}\label{e56}
{\bf n}_w\times {\bf A}_w(u) = \left[\cosh(\lambda\,u)+\alpha\,\sinh(\lambda\,u)\right]\,{\bf n}_w\times{\bf A}_w(0),
\end{equation}
where $\alpha$ is an arbitrary constant. Here, we have made the simplifying assumption that ${\bf A}_w(u)$ does
not change direction across the wall: i.e., as $u$ varies from 0 to 1. This assumption is certainly valid in the thin-shell
limit, and we shall assume that it is also valid in the thick-shell limit. To lowest order in $\delta_\lambda$, 
\begin{equation}\label{e57}
\nabla\times {\bf A}_w(u)\simeq \frac{1}{d}\,{\bf n}_w\times \frac{\partial {\bf A}_w}{\partial u} = 
\frac{\lambda}{d}\left[\sinh(\lambda\,u)+\alpha\,\cosh(\lambda\,u)\right]{\bf n}\times{\bf A}_w(0).
\end{equation}

The boundary conditions that must be satisfied at the inner and outer surfaces of the wall are continuity of the tangential
component of the electric field,\cite{jackson} and continuity of the tangential component of the magnetic field. The
latter boundary condition follows because we are assuming that there are no sheet currents flowing on the inner or the
outer surface of the wall.\cite{hf} Thus, we obtain
\begin{align}
{\bf n}_w\times {\bf A}_i&= {\bf n}_w\times {\bf A}_w(0),\\[0.5ex]
{\bf n}_w\times {\bf A}_o&= {\bf n}_w\times {\bf A}_w(1)= (\cosh\lambda+\alpha\,\sinh\lambda)\,{\bf n}\times {\bf A}_i,\label{e59}\\[0.5ex]
\bar{b}\,{\bf n}_w\times (\nabla \times{\bf A}_i)&= \bar{b}\,[{\bf n}_w\times(\nabla\times{\bf A}_w)](0)\simeq \delta_\lambda^{-1}\,\alpha\,{\bf n}_w\times({\bf n}_w\times
{\bf A}_i),\label{e60}\\[0.5ex]
\bar{b}\,{\bf n}_w\times (\nabla \times{\bf A}_o)&= \bar{b}\,[{\bf n}_w\times(\nabla\times{\bf A}_w)](1)\simeq \delta_\lambda^{-1}\,(\sinh\lambda+ \alpha\cosh\lambda)\,{\bf n}_w\times({\bf n}_w\times
{\bf A}_i)\nonumber\\[0.5ex]
&= \delta_\lambda^{-1}\left(\frac{\sinh\lambda+ \alpha\,\cosh\lambda}{\cosh\lambda+\alpha\sinh\lambda}\right){\bf n}_w\times({\bf n}_w\times
{\bf A}_o),\label{e61}
\end{align}
where use has been made of Eqs.~(\ref{e51}), (\ref{e56}), and (\ref{e57}). Here, ${\bf A}_i$ and ${\bf A}_o$ are evaluated on the inner and
outer surfaces of the wall, respectively. 

Let
\begin{equation}\label{e62}
\zeta_o = \frac{\bar{b}\,{\bf n}_w\times (\nabla \times{\bf A}_o)}
{{\bf n}_w\times({\bf n}_w\times{\bf A}_o)}.
\end{equation}
Note that $\zeta_o$ must be a scalar in order to satisfy Eq.~(\ref{e61}). 
Now, we expect ${\bf A}_o$ to decay monotonically to zero as we move from the outer surface of the wall to infinity. Moreover, we expect the
decay length to be of order $1/\bar{b}$. It follows that $\zeta_o\sim {\cal O}(1)$. 

Equations~(\ref{e61}) and (\ref{e62}) can be combined to give
\begin{equation}
\delta_\lambda\,\zeta_o = \frac{\sinh\lambda+\alpha\,\cosh\lambda}{\cosh\lambda+ \alpha\sinh\lambda}.
\end{equation}
It follows that\,\cite{chap}
\begin{equation}
\alpha =- \tanh\lambda + \frac{\delta_\lambda\,\zeta_o}{\cosh^2\lambda} + {\cal O}(\delta_\lambda\,\zeta_o)^2.
\end{equation}
To lowest order in $\delta_\lambda\,\zeta_o$, Eqs.~(\ref{e59}), (\ref{e60}), and (\ref{e62}) yield 
\begin{align}\label{e66}
{\bf n}_w\times {\bf A}_o&=\frac{{\bf n}_w\times {\bf A}_i}{\cosh\lambda},\\[0.5ex]
{\bf n}_w\times(\nabla\times {\bf A}_i) &=-\frac{\lambda\,\tanh\lambda}{d}\,{\bf n}_w\times({\bf n}_w\times {\bf A}_i) + \frac{{\bf n}_w\times(\nabla\times{\bf A}_o)}{\cosh\lambda}.\label{e67}
\end{align}
The previous equation gives 
\begin{equation}\label{e68}
{\bf n}_w\times[{\bf n}_w\times(\nabla\times {\bf A}_i)] =\frac{\lambda\,\tanh\lambda}{d}\,{\bf n}_w\times {\bf A}_i + \frac{{\bf n}_w\times[{\bf n}_w\times(\nabla\times{\bf A}_o)]}{\cosh\lambda}.
\end{equation}
Equations~(\ref{e66}) and (\ref{e67}) [or, alternatively, Eq.~(\ref{e68})] are the two matching conditions that must be satisfied at the wall. According to Eq.~(\ref{e66}),
$|{\bf n}_w\times {\bf A}_o|\simeq |{\bf n}_w\times {\bf A}_i|$ in the thin-shell limit.\cite{hf} On the other hand,
$|{\bf n}_w\times {\bf A}_o|<|{\bf n}_w\times {\bf A}_i|$ in the thick-shell limit, due to the shielding of the outer vacuum region, $V_o$, from the
inner vacuum region, $V_i$, by eddy currents excited in the wall. Finally, Eqs.~(\ref{e66}) and (\ref{e68}) can be combined to produce
\begin{equation}\label{e69}
{\bf n}_w\times {\bf A}_i\cdot {\bf n}_w\times[{\bf n}_w\times(\nabla\times {\bf A}_i)]= {\bf n}_w\times {\bf A}_o\cdot {\bf n}_w\times[{\bf n}_w\times(\nabla\times {\bf A}_o)]+ \frac{\lambda\,\tanh\lambda}{d}\,|{\bf n}_w\times {\bf A}_i|^2.
\end{equation}

\subsection{Timescale Ordering}
In order for the left-hand side of Eq.~(\ref{e6}) to compete with the right-hand side, we need
$\gamma\sim \tau_A^{-1}$, where $\tau_A \simeq \sqrt{\mu_0\,\rho\,a^2/B^2}\sim 10^{-7}$\,s. Here, $a$ is the minor
radius of the plasma. However, the growth-rate of a resistive wall mode is of order $\bar{\tau}_w\simeq \mu_0\,\sigma_w\,d\,\bar{b}\gtrsim 10^{-3}$\,s.\cite{rwm1,rwm2,hf}
Hence, it is clear that, for the case of a resistive wall mode, the left-hand side of Eq.~(\ref{e6}) is negligible. In other words, plasma inertia
is negligible. Consequently, the plasma displacement associated with the resistive wall mode satisfies the force balance equation, (\ref{e30}),
within the plasma. Note that we are saying that the actual physical solution, as opposed to a trial solution, must satisfy Eq.~(\ref{e30}) within
the plasma. 

\subsection{Self-Adjoint Property of Force Operator}\label{energy}
We wish to prove that the force operator remains self-adjoint in the presence of a resistive wall. The proof follows along the
lines of that outlined in Sect.~\ref{self}, except that
\begin{align}
\int_{S_p}{\bf n}\cdot {\bf C}\times (\nabla\times {\bf A})\,dS_p&
= -\int_{V_i}\nabla\cdot[{\bf C}_i\times (\nabla\times {\bf A}_i)]\,dV_i +\int_{S_w}{\bf n}_w\cdot{\bf C}_i\times(\nabla\times{\bf A}_i)\,dS_w\nonumber\\[0.5ex]
&= -\int_{V_i}(\nabla\times {\bf C}_i)\cdot(\nabla\times {\bf A}_i)\,dV_i\nonumber\\[0.5ex]&\phantom{=}- \int{\bf n}_w\times {\bf C}_i\cdot{\bf n}_w\times[{\bf n}_w\times(\nabla\times{\bf A}_i)]\,dS_w,
\end{align}
where use has been made of Eq.~(\ref{e9}). However, Eq.~(\ref{e69}) can easily be generalized to give 
\begin{align}
{\bf n}_w\times {\bf C}_i\cdot {\bf n}_w\times[{\bf n}_w\times(\nabla\times {\bf A}_i)]&= {\bf n}_w\times {\bf C}_o\cdot {\bf n}_w\times[{\bf n}_w\times(\nabla\times {\bf A}_o)]\nonumber\\[0.5ex]&\phantom{=}+ \frac{\lambda\,\tanh\lambda}{d}\,{\bf n}_w\times {\bf C}_i\cdot{\bf n}_w\times {\bf A}_i.
\end{align}
Moreover,
\begin{align}
\int_{S_w} {\bf n}_w\times {\bf C}_o\cdot {\bf n}_w\times[{\bf n}_w\times(\nabla\times {\bf A}_o)\,dS_w
&= -\int_{S_w} {\bf n}_w\times {\bf C}_o\cdot (\nabla\times {\bf A}_o)\,dS_w\nonumber\\[0.5ex]
&=
\int_{V_o}\nabla\cdot[{\bf C}_o\times (\nabla\times {\bf A}_o)]\,dV_o\nonumber\\[0.5ex]
&= \int_{V_o}(\nabla\times {\bf C}_o)\cdot(\nabla\times {\bf A}_o)\,dV_o,
\end{align}
where use has been made of Eqs.~(\ref{e9}) and (\ref{e14}). Hence, we deduce that
\begin{align}
\int_{S_p}{\bf n}\cdot {\bf C}\times (\nabla\times {\bf A})\,dS_p&
= -\int_{V_i}(\nabla\times {\bf C}_i)\cdot(\nabla\times {\bf A}_i)\,dV_i -\int_{V_o}(\nabla\times {\bf C}_o)\cdot(\nabla\times {\bf A}_o)\,dV_o\nonumber\\[0.5ex]
&\phantom{=}-
\int_{S_w}\frac{\lambda\,\tanh\lambda}{d}\,{\bf n}_w\times {\bf C}_i\cdot{\bf n}_w\times {\bf A}_i\,dS_w.
\end{align}
Thus, in the presence of a resistive wall, Eq.~(\ref{e18}) generalizes to give
\begin{align}
 \int_{V_p} \bta\cdot{\bf F}(\bxi)\,dV_p&=-  \int_{V_p}\left[{\mit\Gamma}\,p\,(\nabla\cdot\bxi)\,(\nabla\cdot\bta)+\mu_0^{-1}\,{\bf Q}\cdot{\bf R}+\frac{1}{2}\,\nabla p\cdot[(\nabla\cdot\bta)\,\bxi
+(\nabla\cdot\bxi)\,\bta]\right.\nonumber\\[0.5ex]
&\phantom{=}\left.+\frac{1}{2}\,{\bf j}\cdot(\bxi\times {\bf R}+\bta\times {\bf Q})\right]d{V_p}\nonumber\\[0.5ex]
&\phantom{=} -\int_{S_p}  ({\bf n}\cdot\bta)\,({\bf n}\cdot\bxi)\,{\bf n}\cdot \left\llbracket\nabla\left(p+ \frac{B^{2}}{2\,\mu_0}\right)\right\rrbracket d{S_p}
\nonumber\\[0.5ex]
&\phantom{=} -\int_{V_i}\mu_0^{-1}(\nabla\times{\bf C}_i)\cdot(\nabla\times {\bf A}_i)\,dV_i -\int_{V_o}\mu_0^{-1}\,(\nabla\times {\bf C}_o)\cdot(\nabla\times {\bf A}_o)\,dV_o\nonumber\\[0.5ex]
&\phantom{=}-
\int_{S_w}\mu_0^{-1}\frac{\lambda\,\tanh\lambda}{d}\,{\bf n}_w\times {\bf C}_i\cdot{\bf n}_w\times {\bf A}_i\,dS_w.
\end{align}
The  self-adjoint property of the force operator, (\ref{e19}), immediately follows from the symmetric nature of the previous equation. Thus, we conclude that the force operator
remains self-adjoint in the presence of a resistive wall, even when the wall lies in the thick-shell limit. 

\subsection{Variational Principle}\label{svar}
Making use of the previous equation, the perturbed potential energy of the resistive wall mode can be written
\begin{equation}\label{e75}
\delta W(\bxi,\bxi) = -\frac{1}{2}\int_{V_p}\bxi\cdot{\bf F}(\bxi)\,dV_p = \delta W_p + \delta W_s + \delta W_v^{(i)}+\delta W_v^{(o)}+\delta W_v^{(w)},
\end{equation}
where
\begin{align}\label{e75}
\delta W_v^{(i)}({\bf A}_i,{\bf A}_i) &= \frac{1}{2\,\mu_0}\int_{V_i}(\nabla\times{\bf A}_i)^2\,dV_i,\\[0.5ex]
\delta W_v^{(o)}({\bf A}_o,{\bf A}_o)&= \frac{1}{2\,\mu_0}\int_{V_o}(\nabla\times{\bf A}_o)^2\,dV_o,\label{e76}\\[0.5ex]
\delta W_v^{(w)}({\bf A}_i,{\bf A}_i) &=\frac{\lambda\,\tanh\lambda}{d}\, \frac{1}{2\,\mu_0}\int_{S_w}|{\bf A}_i\times{\bf n}_w|^2\,dS_w.\label{e77}
\end{align}
Clearly, $\delta W_v^{(i)}$ is the potential energy associated with the vacuum region interior to the wall, $\delta W_v^{(o)}$ is the potential energy associated with the
vacuum region exterior to the wall, and $\delta W_v^{(w)}$ is the potential energy associated with eddy currents excited in the wall. In fact, it is easily
demonstrated that
\begin{equation}
\delta W_v^{(w)} = \gamma^{-1}\int_{V_w}{\bf j}_w\cdot{\bf E}_w\,dV_w,
\end{equation}
where $V_w$ represents the volume occupied by the wall. In deriving the previous formula, we have assumed that ${\bf n}_w\cdot{\bf A}_w\simeq 0$ 
in $V_w$, 
which implies that the eddy currents excited in the wall flow predominately in directions tangental to the wall surface
(because $d\ll \bar{b}$). Incidentally, it is easily demonstrated that the magnetic energy of the wall, 
$(1/2\,\mu_0)\,\int_{V_w}|\nabla\times{\bf A}_w|^2\,dV_w$, is negligible. 

Given that Eq.~(\ref{e30})  holds within
the plasma, it is clear that from Eq.~(\ref{e19a}) that 
\begin{equation}\label{e79}
\delta W= 0
\end{equation}
for a resistive wall mode. This is not a surprising result, because a resistive wall mode is essentially a marginally stable ideal mode (given
that $|\gamma|\ll \tau_A^{-1}$), which corresponds to a mode with $\delta W=0$. 

Consider
\begin{equation}
\delta[\delta W(\bxi,\bxi)] = \delta W(\delta\bxi,\bxi)+ \delta W(\bxi,\delta\bxi) = 2\,\delta W(\delta\bxi,\bxi),
\end{equation}
where use has been made of the self-adjoint property of the force operator, (\ref{e19}). Now, $2\,\delta W_p(\bxi,\bxi)$
and $2\,\delta W_s(\delta\bxi,\bxi)$ are again given by Eqs.~(\ref{e25}) and (\ref{e26}), respectively.  Moreover,
\begin{align}
2\,\mu_0\,\delta W_v^{(i)}(\delta {\bf A}_i,{\bf A}_i) &= \int_{V_i} \nabla\times \delta {\bf A}_i\cdot\nabla\times{\bf A}_i\,dV_i\nonumber\\[0.5ex]
&=\int_{V_i} \left\{\nabla\cdot[\delta {\bf A}_i\times (\nabla\times{\bf A}_i)] + \delta{\bf A}_i\cdot\nabla\times(\nabla\times {\bf A}_i)\right\}dV_i\nonumber\\[0.5ex]
&= -\int_{S_p}{\bf n}\cdot\delta {\bf A}\times(\nabla\times {\bf A})\,dS_p + \int_{S_w}{\bf n}_w\cdot\delta {\bf A}_i\times (\nabla\times {\bf A}_i)\,dS_w
\nonumber\\[0.5ex]&\phantom{=}+\int_{V_i}\delta{\bf A}_i\cdot\nabla\times(\nabla\times {\bf A}_i)\,dV_i\nonumber\\[0.5ex]
&=\int_{S_p}({\bf n}\cdot\bxi)\,\hat{\bf B}\cdot (\nabla\times {\bf A})\,dS_p - \int_{S_w}{\bf n}_w\times {\bf \delta A}_i\cdot {\bf n}_w\times [{\bf n}_w\times(\nabla\times {\bf A}_i)]\,dS_w\nonumber\\[0.5ex]&\phantom{=}+\int_{V_i}\delta{\bf A}_i\cdot\nabla\times(\nabla\times {\bf A}_i)\,dV_i,
\end{align}
where use has been made of the essential boundary condition (\ref{e11}). Furthermore, 
\begin{align}
2\,\mu_0\,\delta W_v^{(o)}(\delta {\bf A}_o,{\bf A}_o) &= \int_{V_o} \nabla\times \delta {\bf A}_o\cdot\nabla\times{\bf A}_o\,dV_o\nonumber\\[0.5ex]
&=\int_{V_o} \left\{\nabla\cdot[\delta {\bf A}_o\times (\nabla\times{\bf A}_o)] + \delta{\bf A}_o\cdot\nabla\times(\nabla\times {\bf A}_o)\right\}dV_o\nonumber\\[0.5ex]
&=- \int_{S_w}{\bf n}_w\cdot\delta {\bf A}_o\times (\nabla\times {\bf A}_o)\,dS_w+\int_{V_o}\delta{\bf A}_o\cdot\nabla\times(\nabla\times {\bf A}_o)\,dV_o\nonumber\\[0.5ex]
& =\int_{S_w}{\bf n}_w\times {\bf \delta A}_i\cdot \frac{{\bf n}_w\times [{\bf n}_w\times(\nabla\times {\bf A}_o)]}{\cosh\lambda}\,dS_w\nonumber\\[0.5ex]&\phantom{=}+\int_{V_o}\delta{\bf A}_i\cdot\nabla\times(\nabla\times {\bf A}_o)\,dV_o,
\end{align}
where use has been made of the essential boundary condition (\ref{e14}), as well as Eq.~(\ref{e66}). 
Finally, 
\begin{equation}
2\,\mu_0\,\delta W_v^{(w)}(\delta {\bf A}_i,{\bf A}_i) = \int_{S_w} \frac{\lambda\,\tanh\lambda}{d}\,{\bf n}_w\times \delta{\bf A}_i\cdot{\bf n}_w\times {\bf A}_i\,dS_w.
\end{equation}
Thus, setting $\delta[\delta W(\bxi,\bxi)]$ to zero, we obtain 
\begin{align}
0=&-\int_{V_p}\delta\bxi\cdot{\bf F}({\bxi})\,d{V_p}\nonumber\\[0.5ex]
&-\int_{S_p}({\bf n}\cdot\delta\bxi)\left[-{\mit\Gamma}\,p\,\nabla\cdot\bxi+\bxi\cdot\nabla\left(\frac{B^{2}}{2\,\mu_0}\right)
+\mu_0^{-1}\,{\bf B}\cdot{\bf Q} \right.\nonumber\\[0.5ex]&\phantom{=}\left.-\bxi\cdot\nabla\left(\frac{\hat{B}^{2}}{2\,\mu_0}\right)-\mu_0^{-1}\,\hat{\bf B}\cdot(\nabla\times{\bf A})\right]d{S_p}
\nonumber\\[0.5ex]
&+ \int_{V_i}\mu_0^{-1}\,\delta{\bf A}_i\cdot\nabla\times (\nabla\times{\bf A}_i)\,d{V_i}+ \int_{V_o}\mu_0^{-1}\delta{\bf A}_o\cdot\nabla\times (\nabla\times{\bf A}_o)\,d{V_o}\nonumber\\[0.5ex]
&+ \mu_0^{-1}\int_{S_w}{\bf n}_w\times\delta {\bf A}_i\cdot\left[\frac{{\bf n}_w\times [{\bf n}_w\times(\nabla\times {\bf A}_o)]}{\cosh\lambda}-
{\bf n}_w\times [{\bf n}_w\times(\nabla\times {\bf A}_i)]\right.\nonumber\\[0.5ex]
&\phantom{=}\left. + \frac{\lambda\,\tanh\lambda}{d}\,{\bf n}_w\times {\bf A}_i\right].
\end{align}
However, the previous equation must hold for arbitrary $\delta\bxi$ and $\delta{\bf A}$. Hence, we deduce that the solution pair, [$\bxi({\bf r})$,
${\bf A}({\bf r})]$, that minimizes the $\delta W(\bxi,\bxi)$ specified in Eq.~(\ref{e75}) satisfies Eq.~(\ref{e30}) in $V_p$, satisfies
Eq.~(\ref{e9}) in $V_i$ and $V_o$, satisfies  the pressure balance matching condition, (\ref{e12}), at the plasma boundary, and
satisfies the matching condition (\ref{e68}) at the wall. In other words, the solution pair solves the resistive wall mode problem. 

Note that the wall matching condition (\ref{e66}) is clearly an essential
boundary condition, whereas the matching condition (\ref{e67}) [or, alternatively, (\ref{e68})] plays the role of a natural boundary condition.\cite{freid1}

We might ask why we believe that the
solution pair that minimizes the $\delta W(\bxi,\bxi)$ specified in Eq.~(\ref{e75}) will satisfy the pressure balance matching condition, (\ref{e12}), at the plasma boundary,
when the solution pair that minimizes the $\delta W(\bxi,\bxi)$  specified in Eq.~(\ref{e21}) failed to do this. The answer is that the unspecified
growth-rate of the resistive wall mode, $\gamma$, introduces an additional degree of freedom into the system that allows us to simultaneously
satisfy all of the boundary conditions. 

\subsection{Generalized Haney-Freidberg Formula}\label{minimize}
Following Haney \& Freidberg,\cite{hf} let us write
\begin{align}\label{e85}
{\bf A}_i({\bf r}) &= c_1\,{\bf A}_{nw}({\bf r}) + c_2\,{\bf A}_{pw}({\bf r}),\\[0.5ex]
{\bf A}_o({\bf r}) &= c_3\,{\bf A}_{nw}({\bf r}),\label{e86}
\end{align}
where $c_1$, $c_2$, and $c_3$ are constants that must be determined by the essential boundary conditions and the minimization process.
Now, the essential boundary condition (\ref{e11}) implies
that
\begin{align}
{\bf n}\times {\bf A}_i = -({\bf n}\cdot\bxi)\,\hat{\bf B}
\end{align}
on $S_p$. Combining this relation with Eqs.~(\ref{e32}), (\ref{e36}), and (\ref{e85}), we deduce that
\begin{equation}\label{e89}
c_1+ c_2= 1.
\end{equation}
Next, the essential boundary condition (\ref{e66}) can be combined with Eqs.~(\ref{e33}), (\ref{e85}), and (\ref{e86})  to give
\begin{equation}
c_3 = \frac{c_1}{\cosh\lambda}.
\end{equation}
Note that the essential boundary condition (\ref{e14}) is automatically satisfied because of Eq.~(\ref{exx}). 

The following results are easily demonstrated:
\begin{align}
2\,\mu_0\,\delta W_v^{(i)}&= \int_{S_p}({\bf n}\cdot\bxi)\,\hat{\bf B}\cdot\nabla\times {\bf A}_i\,dS_p + \int_{S_w}{\bf n}_w\times {\bf A}_i\cdot\nabla\times{\bf A}_i\,dS_w\nonumber\\[0.5ex]
&=c_1\int_{S_p}({\bf n}\cdot\bxi)\,\hat{\bf B}\cdot\nabla\times {\bf A}_{nw}\,dS_p +c_2\int_{S_p}({\bf n}\cdot\bxi)\,\hat{\bf B}\cdot\nabla\times {\bf A}_{pw}\,dS_p 
\nonumber\\[0.5ex]&\phantom{=} +c_1^{\,2} \int_{S_w}{\bf n}_w\times {\bf A}_{nw}\cdot\nabla\times{\bf A}_{nw}\,dS_w+c_1\,c_2 \int_{S_w}{\bf n}_w\times {\bf A}_{nw}\cdot\nabla\times{\bf A}_{pw}\,dS_w,\\[0.5ex]
2\,\mu_0\,\delta W_v^{(o)}&= - \int_{S_w}{\bf n}_w\times {\bf A}_o\cdot\nabla\times{\bf A}_o\,dS_w=- c_3^{\,2}\int_{S_w}{\bf n}_w\times {\bf A}_{nw}\cdot\nabla\times{\bf A}_{nw}\,dS_w,\\[0.5ex]
2\,\mu_0\,\delta W_v^{(b)}&=  \int_{S_p}({\bf n}\cdot\bxi)\,\hat{\bf B}\cdot\nabla\times {\bf A}_{pw}\,dS_p,\\[0.5ex]
2\,\mu_0\,\delta W_v^{(\infty)}&=  \int_{S_p}({\bf n}\cdot\bxi)\,\hat{\bf B}\cdot\nabla\times {\bf A}_{nw}\,dS_p,
\end{align}
and
\begin{align}
\int_{V_i}(\nabla\times{\bf A}_{nw})\cdot(\nabla\times{\bf A}_{pw})\,dV_i &= \int_{S_p}({\bf n}\cdot\bxi)\,\hat{\bf B}\cdot\nabla\times {\bf A}_{pw}\,dS_p 
+\int_{S_w}{\bf n}_w\times {\bf A}_{nw}\cdot\nabla\times{\bf A}_{pw}\,dS_w\nonumber\\[0.5ex]
&=\int_{S_p}({\bf n}\cdot\bxi)\,\hat{\bf B}\cdot\nabla\times {\bf A}_{nw}\,dS_p,
\end{align}
where use has been made of Eqs.~(\ref{e9}), (\ref{e11}), (\ref{e13}), (\ref{e14}), (\ref{e32}), (\ref{e33}), (\ref{e36}), (\ref{exx}),  (\ref{e85}), and (\ref{e86}). 
It is helpful to define
\begin{equation}\label{wx}
2\,\mu_0\,\delta W_v^{(x)} = \int_{V_o} |\nabla\times {\bf A}_{nw}|^2\,dV_p = -\int_{S_w}{\bf n}\times {\bf A}_{nw}\cdot\nabla\times {\bf A}_{nw}\,dS_w.
\end{equation}
Here, $\delta W_v^{(x)}$ represents the contribution of the region $V_o$ to the no-wall vacuum energy. 
The previous eight equations can be combined to the give the following expression for the
vacuum energy:
\begin{align}\label{e96}
\delta W_v &= \delta W_v^{(i)}+\delta W_v^{(o)}+ \delta W_v^{(w)}\nonumber\\[0.5ex]
&=\delta W_v^{(\infty)} + c_2^{\,2}\,(\delta W_v^{(b)}-\delta W_v^{(\infty)})\nonumber\\[0.5ex]
&\phantom{=}+ (1-c_2)^2\left[\frac{\lambda\,\tanh\lambda}{2\,\mu_0\,d}\int_{S_w}
|{\bf A}_{nw}\times {\bf n}_w|^2\,dS_w - \tanh^2\lambda\,\delta W_{v}^{(x)}\right].
\end{align}
Now, the ratio of the first term to the second term appearing in square brackets in the previous equation is
\begin{equation}
\frac{\bar{b}\,\lambda}{d\,\tanh\lambda}.
\end{equation}
Thus, in the thin-shell limit, $\lambda\ll 1$, the ratio is $\bar{b}/d$, which is very large. On the other hand, in the
thick-shell limit, $\lambda\gg 1$, the ratio is $\delta_\lambda^{-1}$ [see Eq.~(\ref{e51})], which is also very large. Hence,
we deduce that we can neglect the second term. Thus,  the previous equation simplifies to give
\begin{align}
\delta W_v &=\delta W_v^{(\infty)} + c_2^{\,2}\,(\delta W_v^{(b)}-\delta W_v^{(\infty)})+ (1-c_2)^2\,F(\lambda),
\end{align}
where
\begin{equation}\label{e100}
F(\lambda)= \frac{\lambda\,\tanh\lambda}{2\,\mu_0\,d}\int_{S_w}
|{\bf A}_{nw}\times {\bf n}_w|^2\,dS_w.
\end{equation}

According to the variational principle proved in Sect.~\ref{svar}, we can determine the true vacuum energy by minimizing  
$\delta W_v$ with respect to variations in $c_2$. This procedure
yields
\begin{align}\label{e101x}
c_2 &= \frac{F(\lambda)}{\delta W_v^{(b)}-\delta W_v^{(\infty)}+F(\lambda)},\\[0.5ex]
\delta W_v& = \delta W_v^{(\infty)} + \frac{(\delta W_v^{(b)}-\delta W_v^{(\infty)})\,F(\lambda)}{\delta W_v^{(b)}-\delta W_v^{(\infty)}+F(\lambda)}.
\end{align}
Following Haney \& Freidberg,\cite{hf} we can define the effective minor radius of the wall as
\begin{equation}\label{e101}
\bar{b} = \frac{(1/2\,\mu_0)\int_{S_w}
|{\bf A}_{nw}\times {\bf n}_w|^2\,dS_w}{\delta W_v^{(b)}-\delta W_v^{(\infty)}}.
\end{equation}
Equations (\ref{e100}) and (\ref{e101}) give
\begin{equation}\label{e102}
\delta W_v = \delta W_v^{(\infty)} + (\delta W_v^{(b)}-\delta W_v^{(\infty)})\,\frac{G(\gamma)}{1+G(\gamma)},
\end{equation}
where
\begin{align}
G(\gamma) &= \sqrt{\frac{\gamma\,\bar{\tau}_w}{\bar{\delta}_w}}\,\tanh\left(\!\sqrt{\bar{\delta}_w\,\gamma\,\bar{\tau}_w}\right),\\[0.5ex]
\bar{\tau}_w &= \mu_0\,\sigma_w\,d\,\bar{b},\label{bart}\\[0.5ex]
\bar{\delta}_w& = \frac{d}{\bar{b}}\label{delt}.
\end{align}
Here, $\bar{\tau}_w$ is the effective L/R time of the wall, whereas $\bar{\delta}_w\ll 1$ measures the relative wall thickness. 

According to Eqs.~(\ref{e75}), (\ref{e96}), and (\ref{e102}), the total potential energy of the perturbation is
\begin{align}\label{e108}
\delta W &= \delta W_p+\delta W_s + \delta W_v
= \delta W_{nw} +  (\delta W_{pw}-\delta W_{nw})\,\frac{G(\gamma)}{1+G(\gamma)},
\end{align}
where use has been made of Eqs.~(\ref{e30s}) and (\ref{e34}). Finally, Eq.~(\ref{e79}) mandates that $\delta W=0$ for a resistive wall mode, so we obtain the following {\em generalized Haney-Freidberg formula}:
\begin{equation}\label{hf1}
 \sqrt{\frac{\gamma\,\bar{\tau}_w}{\bar{\delta}_w}}\,\tanh\left(\!\sqrt{\bar{\delta}_w\,\gamma\,\bar{\tau}_w}\right)= - \frac{\delta W_{nw}}{\delta W_{pw}}
\end{equation}
for the case $\delta W_{nw}<0$, $\delta W_{pw}>0$ in which the resistive wall mode is unstable, and
\begin{equation}\label{hf2}
 \sqrt{\frac{-\gamma\,\bar{\tau}_w}{\epsilon_w}}\,\tan\left(\!\sqrt{-\bar{\delta}_w\,\gamma\,\bar{\tau}_w}\right)=  \frac{\delta W_{nw}}{\delta W_{pw}}
\end{equation}
for the case $\delta W_{nw}>0$, $\delta W_{pw}>0$ in which the resistive wall mode is stable. The formula does not
apply to the case $\delta W_{pw}<0$ in which the plasma is ideally unstable in the presence of the wall, because the neglect
of plasma inertia is not tenable in this  situation. The derivation of the generalized Haney-Freidberg formula
is valid provided
\begin{equation}
\frac{d}{\bar{b}}\ll |\gamma|\,\bar{\tau}_w, ~1.
\end{equation}
In the thin-shell limit, $\epsilon_w\,|\gamma|\,\tau_w \ll 1$, Eqs.~(\ref{hf1})--(\ref{hf2}) reduce to the original Haney-Freidberg formula:
\begin{equation}
\gamma\,\bar{\tau}_w = - \frac{\delta W_{nw}}{\delta W_{pw}}.
\end{equation}

Equations~(\ref{e79}), (\ref{e89}), (\ref{e101x}), and (\ref{e108}) yield
\begin{align}\label{c1}
c_1 &= \frac{\delta W_{pw}}{\delta W_{pw}-\delta W_{nw}},\\[0.5ex]
c_2&= -\frac{\delta W_{nw}}{\delta W_{pw}-\delta W_{nw}}.\label{c2}
\end{align}
Moreover, given that the gereralized Haney-Freidberg dispersion relation sets $\delta W=0$, a comparison of Eqs.~(\ref{e38}) and (\ref{e108})
reveals that $\delta W_c=0$, where $\delta W_c$ is defined in Eq.~(\ref{e39}).  Thus, it is plausible that the linear combination of solutions, (\ref{e85}), with $c_1$ and $c_2$ given by the
previous two equations, ensures that the pressure balance matching condition, (\ref{e12}), is satisfied at the plasma boundary. Later on, 
in Sect.~\ref{match}, we shall
show this explicitly. 

\subsection{Resistive Wall Mode Growth-Rate}\label{growth}
\begin{figure}
\centerline{\includegraphics[width=\textwidth]{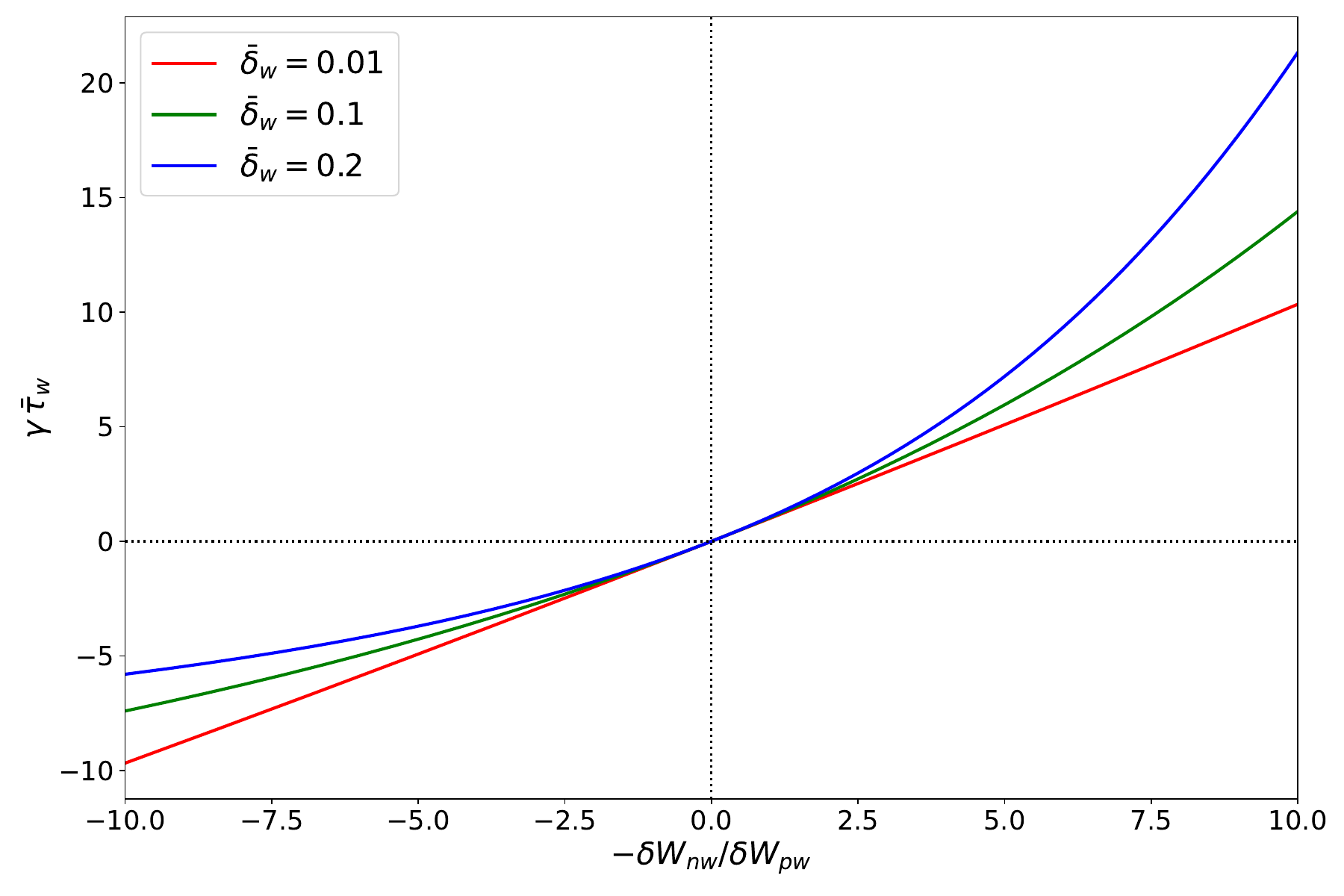}}
\caption{Growth-rate of the resistive wall mode predicted by the  generalized Haney-Freidberg formula, (\ref{hf1}).}\label{fig2}
\end{figure}

Figure~\ref{fig2} shows the growth-rate of the resistive wall mode predicted by the generalized Haney-Freidberg formula, (\ref{hf1})--(\ref{hf2}). 
For a thin wall  characterized by $\bar{\delta}_w\ll 1$, we reproduce the characteristic linear relationship between $\gamma\,\bar{\tau}_w$ and $-\delta W_{nw}/\delta W_{pw}$
predicted by the original Haney-Freidberg formula.
The mode grows or decays on the characteristic L/R timescale, $\bar{\tau}_w$, and the marginal stability point, $\delta W_{nw}=0$, is the
same as that for an ideal-kink mode in the absence of a wall. Thick-wall effects, which manifest themselves when $\bar{\delta}_w\lesssim 1$,  do not
change the marginal stability point, but introduce an interesting asymmetry between growing and decaying modes. Growing modes have
growth-rates that exceed those predicted by the original Haney-Freidberg formula. (Here, we are comparing thick and thin walls with the
same $\bar{\tau}_w$ values.) On the other hand, decaying modes have decay-rates that are less than those predicted by the original 
formula. Note that there are actually multiple branches of decaying solutions, and we have plotted the most slowly decaying branch
in Fig.~\ref{fig2}. 

For a very rapidly growing resistive wall mode, such that $\gamma\,\bar{\tau}_w\,\bar{\delta}_w\gg 1$, which corresponds to the complete
breakdown of the thin-shell approximation, Eq.~(\ref{hf1}) reduces
to\,\cite{thick0, thick2}
\begin{equation}
\gamma\,\bar{\tau}_w = \bar{\delta}_w \left(- \frac{\delta W_{nw}}{\delta W_{pw}}\right)^2.
\end{equation}
In this limit, the mode is almost completely shielded from the vacuum region outside the wall. [See Eq.~(\ref{e66}).]
According to Eq.~(\ref{hf2}), a very rapidly
decaying resistive wall mode (on the slowest decaying solution branch) cannot decay faster than
\begin{equation}
-\gamma\,\bar{\tau}_w = \frac{\pi}{2\,\bar{\delta}_w}.
\end{equation}
Interestingly, rapidly growing resistive wall modes only partially penetrate the wall (in other words, the skin-depth in the wall material is much
less than the wall thickness), whereas rapidly decaying modes always penetrate the wall  (in other words, the skin-depth in the wall material is of order the wall thickness).

\subsection{Further Generalization of Haney-Freidberg Formula}
Our analysis has assumed that the thickness, $d$, and conductivity, $\sigma_w$, of the wall are uniform. However, if we redo the previous analysis
 without making this assumption then very similar arguments reveal that 
\begin{equation}\label{general}
\frac{\int_{S_w} \! \!\sqrt{\gamma\,\bar{\tau}_w/\bar{\delta}_w}\tanh\left(\!\sqrt{\bar{\delta}_w\,\gamma\,\bar{\tau}_w}\right)|{\bf A}_{nw}\times {\bf n}_w|^2\,dS_{w}}
{\int_{S_w} |{\bf A}_{nw}\times {\bf n}_w|^2\,dS_{w}}=-\frac{\delta W_{nw}}{\delta W_{pw}},
\end{equation}
where both $d$ and $\sigma_w$, which appear inside the expressions for $\bar{\tau}_w$ and $\bar{\delta}_w$ [see Eqs.~(\ref{bart}) and (\ref{delt})],  are now allowed to vary around the wall. 

\section{Axisymmetric Quasi-Cylindrical Equilibrium}\label{scyl}

\subsection{Introduction}
In order to further illustrate some of the arguments presented in this paper, let us calculate the resistive wall stability of an axisymmetric
toroidal plasma of major radius $R_0$ that is modeled as a periodic cylinder. Let
$r$, $\theta$, $z$ be right-handed cylindrical coordinates. Let the magnetic axis of the plasma corresponds to $r=0$, and
let the cylinder be periodic in the $z$-direction, with periodicity length $2\pi\,R_0$. 

\subsection{Plasma Equilibrium}
The plasma equilibrium is such that $p=p(r)$ and ${\bf B}= B_\theta(r)\,{\bf e}_\theta+B_z(r)\,{\bf e}_z$. 
Force balance within the plasma yields
\begin{equation}
\mu_0\,p' + B_\theta\,B_\theta' + B_z\,B_z' + \frac{B_\theta^{\,2}}{r}=0,
\end{equation}
where $'\equiv d/dr$. Force balance across the plasma boundary, which lies at $r=a$, demands that
\begin{equation}
2\,\mu_0\,p + B_\theta^{\,2}+B_z^{\,2} = \hat{B}_\theta^{\,2} + \hat{B}_z^{\,2}
\end{equation}
at $r=a$. In the vacuum region, $r>a$, we have $\hat{B}_\theta'=-\hat{B}_\theta/r$ and $\hat{B}_z'=0$. 

\subsection{Perturbation}
Let us assume that all perturbed quantities vary with $\theta$ and $z$ as $\exp[\,{\rm i}\,(m\,\theta+k\,z)]$, where $m$ is the poloidal mode number of the instability,
$k=n/R_0$, and $n>0$ is the toroidal mode number. The plasma displacement is written
\begin{equation}
\bxi = \bxi_\perp + \xi_\parallel\,{\bf b} = \xi(r)\,{\bf e}_r + \eta(r)\,{\bf e}_\eta+ \xi_\parallel(r)\,{\bf b},
\end{equation}
where ${\bf e}_\eta = (B_z/B)\,{\bf e}_\theta-(B_\theta/B)\,{\bf e}_z$, and ${\bf b} =  (B_\theta/B)\,{\bf e}_\theta+(B_zB)\,{\bf e}_z$. Here, $\xi$ is assumed to be real,
whereas $\eta$ and $\xi_\parallel$ turn out to be imaginary. 
Note that ${\bf e}_r$, ${\bf e}_\eta$, and ${\bf b}$ form a right-handed set of mutually orthogonal unit vectors. 

\subsection{Plasma Potential Energy}
The perturbed plasma potential energy can be written\,\cite{freid,freid1,goed}
\begin{align}
\delta W_p &= \frac{1}{2\,\mu_0}
\left[\int_0^a[{\mit\Gamma}\,\mu_0\,p\,(\nabla\cdot\bxi^\ast)\,(\nabla\cdot\bxi)+{\bf Q}^\ast\cdot{\bf Q} + (\nabla\cdot\bxi_\perp^\ast)\,[\bxi_\perp\cdot\nabla(\mu_0\,p)]\right.\nonumber\\[0.5ex]&\phantom{=}
\left.+\mu_0\,{\bf j}\cdot\bxi_\perp^\ast\times {\bf Q}\right]\frac{1}{2}\,2\pi\,r\,2\pi\,R_0\,dr,
\end{align}
where $\mu_0\,{\bf j}=\nabla\times{\bf B}$, ${\bf Q}=\nabla\times (\bxi_\perp\times{\bf B})$, and the factor $1/2$ comes from averaging $\cos^2(m\,\theta+k\,z)$. After a great deal of
standard analysis, we arrive at 
\begin{equation}
\delta W_p = \frac{\pi^2\,R_0}{\mu_0}\int_0^a W(r)\,r\,dr,
\end{equation}
where
\begin{align}\label{e122}
W(r)&={\mit\Gamma}\,\mu_0\,p\left|
\frac{(r\,\xi)'}{r} + {\rm i}\,\frac{G}{B}\,\eta +{\rm i}\,\frac{F}{B}\,\xi_\parallel\right|^2+
\left|\frac{G}{k_0}\,\frac{(r\,\xi)'}{r}+\frac{2\,k\,B_\theta}{r\,k_0}\,\xi+{\rm i}\,k_0\,B\,\eta\right|^2\nonumber\\[0.5ex]
&+A_1\,\xi^{\prime\,2}+A_2\,\xi'\,\xi + A_3\,\xi^2,
\end{align}
and 
\begin{align}
A_1(r)&= \frac{F^2}{k_0^{\,2}},\\[0.5ex]
A_2(r) &= \frac{(k^2\,r^2\,B_z^{\,2}-m^2\,B_\theta^{\,2})}{r^3\,k_0^{\,2}},\\[0.5ex]
A_3(r) &= F^2 + \frac{1}{r^2}\left[B_z^{\,2} -B_\theta^{\,2}-2\,r\,B_\theta'\,B_\theta- \frac{(r\,G^{\,2} + 4\,m\,k\,B_\theta\,B_z)}{r\,k_0^{\,2}}\right],\\[0.5ex]
F(r) &= \frac{m}{r}\,B_\theta + k\,B_z,\\[0.5ex]
G(r) &= \frac{m}{r}\,B_z-k\,B_\theta,\\[0.5ex]
k_0^{\,2}(r)&= \frac{m^2}{r^2}+ k^2. 
\end{align}
Because $\xi_\parallel$ and $\eta$ only appear in Eq.~(\ref{e122}) inside positive-definite terms, we can minimize $\delta W_p$ by choosing
$\xi_\parallel$ and $\eta$ in such a manner as to set these terms to zero. After doing this, and after integrating by parts, we
arrive at\,\cite{freid,freid1,goed,new}
\begin{equation}\label{e129f}
\delta W_p = \frac{\pi^2\,R_0}{\mu_0}\left\{\int_0^a(f\,\xi^{\prime\,2}+g\,\xi^2)\,dr + \left(\frac{k^2\,r^2\,B_z^{\,2}-m^2\,B_\theta^2}{k_0^{\,2}\,r^2}\right)_{a}\,\xi^2(a)\right\},
\end{equation}
where
\begin{align}
f(r) &= \frac{r\,F^2}{k_0^{\,2}},\\[0.5ex]
g(r) &= \frac{2\,k^2}{k_0^{\,2}}\,\mu_0\,p' +\left(\frac{k_0^{\,2}\,r^2-1}{k_0^{\,2}\,r^2}\right)r\,F^{\,2} + \frac{2\,k^2}{r\,k_0^{\,4}}\left(k\,B_z
-\frac{m}{r}\,B_\theta\right)F.
\end{align}

\subsection{Surface Potential Energy}
The perturbed potential energy associated with equilibrium surface currents can easily be shown to take the form
\begin{equation}\label{esurf}
\delta W_s =\frac{\pi^2\,R_0}{\mu_0}\,(B_\theta^{\,2}-\hat{B}_\theta^{\,2})_{a}\,\xi^2(a).
\end{equation}

\subsection{Vacuum Region}
We shall write the divergence- and curl-free perturbed magnetic field in the vacuum region as
\begin{equation}\label{e133}
\hat{\bf Q} = \nabla\times {\bf A} = {\rm i}\,\nabla V.
\end{equation}
Here, $V=V(r)\,\exp[\,{\rm i}\,(m\,\theta+k\,z)]$, where $V(r)$ is assumed to be real. 
Now, the wall matching condition (\ref{e66}) implies that $(m/r)\,A_\theta + k\,A_z=0$. Hence, we deduce that
\begin{align}
{\rm i} \,A_r &= -\frac{m\,k}{r^2\,k_0^{\,4}}\,\frac{dV}{dr},\\[0.5ex]
A_\theta &=-\frac{k}{k_0^{\,2}}\,\frac{dV}{dr},\label{e135s}\\[0.5ex]
A_z&= \frac{m}{r\,k_0^{\,2}}\,\frac{dV}{dr}.\label{e136s}
\end{align}

\subsection{Matching Conditions at Plasma Boundary}
Given that ${\bf n}={\bf e}_r$, the essential boundary condition (\ref{e11}) yields
\begin{equation}\label{e142x}
\hat{F}\,\xi = \frac{dV}{dr},
\end{equation}
at $r=a$, where $\hat{F}(r)= (m/r)\,\hat{B}_\theta+k\,\hat{B}_z$. Moreover, the pressure balance matching condition, (\ref{e12}), reduces to
\begin{equation}\label{e141}
f\,\xi' + \left(\frac{k^2\,r^2\,B_z^{\,2}-m^2\,B_\theta^2}{k_0^{\,2}\,r^2}\right)\xi + (B_\theta^{\,2}-\hat{B}_\theta^{\,2})\,\xi = r\,\hat{F}\,V
\end{equation}
at $r=a$. 

\subsection{Matching Conditions at Wall}
The wall is assumed to be uniform, and located at minor radius $r=b$. 
Given that ${\bf n}_w={\bf e}_r$, the essential boundary condition (\ref{e66}) gives
\begin{equation}\label{e142}
\left(\frac{dV}{dr}\right)_{b_-} =\cosh\lambda\left(\frac{dV}{dr}\right)_{b_+}.
\end{equation}
Moreover, the natural boundary condition (\ref{e67}) reduces to
\begin{equation}\label{e143}
V(b_-) + \frac{\lambda\,\tanh\lambda}{d}\,\frac{1}{k_b^{\,2}}\,\frac{dV(b_-)}{dr} = \frac{V(b_+)}{\cosh\lambda}.
\end{equation}
Here, $b_-$ and $b_+$ denote the inner and outer wall radii, respectively, where it is assumed that the wall is radially thin. Furthermore, $k_b^{\,2}= m^2/b^2+k^2$. 

\subsection{Vacuum Potential Energy}
The contribution of the vacuum region lying internal to the wall to the overall
potential energy of the perturbation can be shown to be [see Eqs.~(\ref{e75}) and (\ref{e133})]
\begin{equation}\label{e137}
\delta W_v^{(i)} = \frac{\pi^2\,R_0}{\mu_0}\left[
-\int_a^{b_-} V\,\nabla^2 V\,r\,dr-(r\,\hat{F}\,\xi\,V)_{a} + \left(r\,\frac{dV}{dr}\,V\right)_{b_-}\right],
\end{equation}
where use has been made of the essential boundary condition (\ref{e142x}). 
Similarly, the contribution of the vacuum region lying outside the wall is [see Eqs.~(\ref{e76}) and (\ref{e133})]
\begin{equation}
\delta W_v^{(o)} = \frac{\pi^2\,R_0}{\mu_0}\left[
- \left(r\,\frac{dV}{dr}\,V\right)_{b_+}-\int_{b_+}^\infty V\,\nabla^2 V\,r\,dr\right].
\end{equation}

\subsection{Wall  Potential Energy}
The contribution of the wall to the overall 
potential energy of the perturbation can be shown to take the form [see Eqs.~(\ref{e77}), (\ref{e135s}), and (\ref{e136s})]
\begin{equation}
\delta W_v^{(w)}= \frac{\pi^2\,R_0}{\mu_0} \,\frac{\lambda\,\tanh\lambda}{d}\,\frac{b}{k_b^{\,2}}\left(\frac{dV}{dr}\right)^2_{b_-}.
\end{equation}

\subsection{Variation Principle}
The total potential energy of the perturbation is
\begin{align}
\delta W &=\delta W_p + \delta W_s + \delta W_v^{(i)} + \delta W_w+ \delta W_v^{(o)}\nonumber\\[0.5ex]
&=
\frac{\pi^2\,R_0}{\mu_0}\left\{\int_0^a (f\,\xi^{\prime\,2}+g\,\xi^2)\,dr+\left[\left(\frac{k^2\,r^2\,B_z^{\,2}-m^2\,B_\theta^2}{k_0^{\,2}\,r^2}\right)\xi^2
+(B_\theta^{\,2}-\hat{B}_\theta^{\,2})\,\xi^2-r\,\hat{F}\,\xi\,V\right]_{a}\right.\nonumber\\[0.5ex]
&\phantom{=}-\int_a^{b_-}V\,\nabla^2 V\,r\,dr +\left(r\,\frac{dV}{dr}\right)_{b_-}\left[V(b_-)
+\frac{\lambda\,\tanh\lambda}{d}\,
\frac{1}{k_b^{\,2}}\left(\frac{dV}{dr}\right)_{b_-}-\frac{V(b_+)}{\cosh\lambda}\right]\nonumber\\[0.5ex]
&\left.\phantom{=}-\int_{b_+}^{\infty}V\,\nabla^2 V\,r\,dr\right\}.
\end{align}
Here, use has been made of the essential boundary condition (\ref{e142}). 
If we  minimize the potential energy, making use of the previously demonstrated  self-adjoint property of the overall expression, then we get
\begin{align}
\delta[\delta W] &=\frac{2\,\pi^2\,R_0}{\mu_0}\left\{\int_0^a \delta\xi\,[-(f\,\xi')'+g\,\xi]\,dr+\delta\xi(a)\left[f\,\xi'+\left(\frac{k^2\,r^2\,B_z^{\,2}-m^2\,B_\theta^2}{k_0^{\,2}\,r^2}\right)\xi
+(B_\theta^{\,2}-\hat{B}_\theta^{\,2})\,\xi\right.\right.\nonumber\\[0.5ex]
&\left.\phantom{=}-r\,\hat{F}\,V\right]_{a}-\int_a^{b_-}\delta V\,\nabla^2 V\,r\,dr +\left(r\,\frac{d\delta V}{dr}\right)_{b_-}\left[V(b_-)
+\frac{\lambda\,\tanh\lambda}{d}\,
\frac{1}{k_b^{\,2}}\left(\frac{dV}{dr}\right)_{b_-}-\frac{V(b_+)}{\cosh\lambda}\right]\nonumber\\[0.5ex]
&\left.\phantom{=}-\int_{b_+}^{\infty}\delta V\,\nabla^2 V\,r\,dr\right\} =0.
\end{align}
Given that the previous expression must hold for arbitrary $\delta \xi$ and $\delta V$, we deduce that the perturbation that minimizes $\delta W$ satisfies
{\em Newcomb's equation},\cite{new}
\begin{equation}\label{e146}
(f\,\xi')'-g\,\xi=0,
\end{equation}
in the plasma,
satisfies 
\begin{equation}\label{e147}
\nabla^2 V\equiv \frac{1}{r}\,\frac{d}{dr}\!\left(r\,\frac{dV}{dr}\right)-\left(\frac{m^2}{r^2}+k^2\right)V=0
\end{equation}
in the vacuum, and satisfies the natural boundary conditions (\ref{e141}) and (\ref{e143}) at the plasma boundary and at the wall, respectively. 

\subsection{No-Wall and Perfect-Wall Energies}\label{sno}
The independent solutions of Eq.~(\ref{e147}) are $I_{|m|}(k\,r)$ and $K_{|m|}(k\,r)$, where $I_m(z)$ and $K_m(z)$ are modified Bessel functions. The
vacuum scalar potential associated with the no-wall solution must satisfy
\begin{equation}
\frac{dV_{nw}}{dr} = \hat{F}(a)\,\xi(a)
\end{equation}
at the plasma boundary [see Eq.~(\ref{e142x})], and
\begin{equation}\label{e149}
V_{nw}=0
\end{equation}
at infinity. [See Eq.~(\ref{exx}).]  It follows that
\begin{equation}\label{e150}
V_{nw}(r)= \frac{\hat{F}(a)\,\xi(a)}{k}\,\frac{K_{|m|}(k\,r)}{K_{|m|}'(k\,a)}.
\end{equation}
Here, $'$ denotes differentiation with respect to argument. 
The vacuum potential energy associated with the no-wall solution is
\begin{equation}\label{e151}
\delta W_v^{(\infty)} = -\frac{\pi^2\,R_0}{\mu_0}\,(r\,\hat{F}\,\xi\,V_{nw})_a,
\end{equation}
where use has been made of Eqs.~(\ref{e75}), (\ref{e133}), (\ref{e142x}), (\ref{e147}), and (\ref{e149}). 

The
vacuum scalar potential associated with the perfect-wall solution must satisfy 
\begin{equation}
\frac{dV_{pw}}{dr} = \hat{F}(a)\,\xi(a)
\end{equation}
at the plasma boundary [see Eq.~(\ref{e142x})], and
\begin{equation}\label{e153}
\frac{dV_{pw}}{dr}=0
\end{equation}
at the wall. [See Eq.~(\ref{e33}).] It follows that 
\begin{equation}\label{e154}
V_{pw}(r)= \frac{\hat{F}(a)\,\xi(a)}{k}\,\frac{I_{|m|}(k\,r)\,K_{|m|}'(k\,b)-K_{|m|}(k\,r)\,I_{|m|}'(k\,b)}{I_{|m|}'(k\,a)\,K_{|m|}'(k\,b)-K_{|m|}'(k\,a)\,I_{|m|}'(k\,b)}.
\end{equation}
The vacuum potential energy associated with the perfect-wall solution is
\begin{equation}\label{e155}
\delta W_v^{(b)} = -\frac{\pi^2\,R_0}{\mu_0}\,(r\,\hat{F}\,\xi\,V_{pw})_a,
\end{equation}
where use has been made of Eqs.~(\ref{e75}), (\ref{e133}), (\ref{e142x}), (\ref{e147}), and (\ref{e153}). 

Now, integrating by parts, and making use of Eq.~(\ref{e146}), the expression, (\ref{e129f}), for the plasma potential energy becomes
\begin{equation}\label{e156}
\delta W_p = \frac{\pi^2\,R_0}{\mu_0}\left\{\xi\left[f\,\xi' +  \left(\frac{k^2\,r^2\,B_z^{\,2}-m^2\,B_\theta^2}{k_0^{\,2}\,r^2}\right)\xi\right]\right\}_a.
\end{equation}
Thus, the total potential energy of the no-wall ideal external-kink mode takes the form 
\begin{align}\label{e157}
\delta W_{nw} &= \delta W_p +\delta W_s+\delta W_{v}^{(\infty)} \nonumber\\[0.5ex]
&= \frac{\pi^2\,R_0}{\mu_0}\left\{\xi\left[f\,\xi' +  \left(\frac{k^2\,r^2\,B_z^{\,2}-m^2\,B_\theta^2}{k_0^{\,2}\,r^2}\right)\xi
+(B_\theta^2-\hat{B}_\theta^{\,2})\,\xi-r\,\hat{F}\,V_{nw}\right]\right\}_a.
\end{align}
Likewise, the  total potential energy of the perfect-wall ideal external-kink mode can be written 
\begin{align}\label{e158}
\delta W_{pw} &= \delta W_p +\delta W_s+\delta W_{v}^{(b)} \nonumber\\[0.5ex]
&= \frac{\pi^2\,R_0}{\mu_0}\left\{\xi\left[f\,\xi' +  \left(\frac{k^2\,r^2\,B_z^{\,2}-m^2\,B_\theta^2}{k_0^{\,2}\,r^2}\right)\xi
+(B_\theta^2-\hat{B}_\theta^{\,2})\,\xi-r\,\hat{F}\,V_{pw}\right]\right\}_a.
\end{align}
In accordance with the discussion in Sect.~\ref{s1d}, we can see that the only reason that the
energies $\delta W_{nw}$ and $\delta W_{pw}$ can take non-zero values is because the solution pairs, $[\xi(r)$, $V(r)]$, from which
they are constructed, do not satisfy the pressure balance matching condition, (\ref{e141}), at the plasma boundary. 

\subsection{Effective Wall Radius}
If follows from Eqs.~(\ref{e150}), (\ref{e151}), (\ref{e154}), and (\ref{e155})  that
\begin{equation}
\delta W_v^{(b)} - \delta W_v^{(\infty)}= \frac{\pi^2\,R_0}{\mu_0}\,\frac{(r\,\hat{F}\,\xi)^2_{a}\,K_{|m|}'(k\,b)}{(k\,a)^2\,K_{|m|}'(k\,a)\,
\left[I_{|m|}'(k\,a)\,K_{|m|}'(k\,b)-K_{|m|}'(k\,a)\,I_{|m|}'(k\,b)\right]}.
\end{equation}
Moreover,
\begin{align}
\frac{1}{2\,\mu_0}\int_{S_w}|{\bf A}_{nw}\times {\bf e}_r|^2\,dS_w &= \frac{\pi^2\,R_0}{\mu_0}\,\frac{b}{k_b^{\,2}}\left(\frac{dV_{nw}}{dr}\right)_b^2\nonumber\\[0.5ex]
&=\frac{\pi^2\,R_0}{\mu_0}\,(r\,\hat{F}\,\xi)^2_{a}\,\frac{b}{(k_b\,a)^2}\left[\frac{K_{|m|}'(k\,b)}{K_{|m|}'(k\,a)}\right]^2,
\end{align}
where use has been made of Eqs.~(\ref{e135s}), (\ref{e136s}), and (\ref{e150}). 
Hence, it follows from Eq.~(\ref{e101}) that the ratio of the effective wall minor radius to the actual wall minor radius is\,\cite{freid1}
\begin{equation}
\frac{\bar{b}}{b} = \frac{(k\,b)^2}{m^2+(k\,b)^2}\,\frac{K_{|m|}'(k\,b)}{K_{|m|}'(k\,a)}\,\left[I_{|m|}'(k\,a)\,K_{|m|}'(k\,b)-K_{|m|}'(k\,a)\,I_{|m|}'(k\,b)\right].
\end{equation}

\subsection{Pressure Balance Matching Condition}\label{match}
Let
\begin{equation}\label{e163}
P_a= \left[f\,\xi' +  \left(\frac{k^2\,r^2\,B_z^{\,2}-m^2\,B_\theta^2}{k_0^{\,2}\,r^2}\right)\xi
+(B_\theta^2-\hat{B}_\theta^{\,2})\,\xi\right]_{a}.
\end{equation}
According to Eqs.~(\ref{c1}), (\ref{c2}), (\ref{e157}), and (\ref{e158}), 
\begin{align}\label{e164}
c_1 &= \frac{[P_a - (r\,\hat{F}\,V_{pw})_a]}{[r\,\hat{F}\,(V_{nw}-V_{pw})]_a},\\[0.5ex]
c_2 &=- \frac{[P_a - (r\,\hat{F}\,V_{nw})_a]}{[r\,\hat{F}\,(V_{nw}-V_{pw})]_a}.
\end{align}
Thus, the linear combination of solutions that satisfies the resistive wall mode problem is
characterized by [see Eq.~(\ref{e85})] 
\begin{equation}
(r\,\hat{F}\,V)_a = c_1\,(r\,\hat{F}\,V_{nw})_a + c_2\,(r\,\hat{F}\,V_{pw})_a = P_a.
\end{equation}
However, the pressure balance matching condition at the plasma boundary, (\ref{e141}), can be written
\begin{equation}
P_a = (r\,\hat{F}\,V)_a.
\end{equation}
Thus, it is clear that the matching condition is satisfied when the growth-rate of the resistive wall mode takes the correct value,
which is specified by [see Eqs.~(\ref{bart}), (\ref{delt}), (\ref{hf1}), and (\ref{hf2})],  
\begin{equation}
 \sqrt{\frac{\gamma\,\tau_w}{\delta_w}}\,\tanh\left(\!\sqrt{\delta_w\,\gamma\,\tau_w}\right)= - \frac{b}{\bar{b}}\,\frac{\delta W_{nw}}{\delta W_{pw}}.
\end{equation}
Here, $\tau_w=\mu_0\,\sigma_w\,d\,b$ is the true time-constant of the wall, and $\delta_w=d/b$ is a measure of its true thickness. 

\subsection{Force-Free Reversed-Field Pinch Equilibrium}
Consider a reversed-field pinch\,\cite{freid1} plasma equilibrium. Let us assume, for the sake of simplicity, that the  equilibrium pressure is negligible. 
In this case, the equilibrium magnetic field, both in the plasma and in the vacuum, satisfies\,\cite{slinky} 
\begin{align}
B_\theta' &= \sigma\,B_z-\frac{B_\theta}{r},\\[0.5ex]
B_z'&= -\sigma\,B_\theta,
\end{align}
where $\sigma(r)=\mu_0\,{\bf j}\cdot{\bf B}/B^2$. Let us adopt the following model $\sigma(r)$ profile,\cite{slinky}
\begin{equation}
\sigma(r)= \left\{\begin{array}{lcc}
\sigma_0\,[1-(r/a)^\alpha]^\nu&~~~&r\leq a\\[0.5ex]
0&&r>a
\end{array}
\right.,
\end{equation}
where $\alpha$, $\nu>0$. 
Note that $\sigma(a)=0$, which implies that there is no equilibrium current sheet flowing around  the plasma boundary. 

The resistive wall mode perturbation can be specified, both in the plasma and in the vacuum,  in terms of the perturbed poloidal flux, $\psi(r)$,\cite{fit} where
\begin{equation}
\psi(r)= \left\{\begin{array}{lcc}
r\,F\,\xi&~~~&r\leq a\\[0.5ex]
r\,(dV/dr)&&r>a
\end{array}
\right..
\end{equation}
The matching condition (\ref{e142x}) becomes
\begin{equation}\label{e173}
\psi(a_-) = \psi(a_+),
\end{equation}
whereas the pressure balance  matching condition, (\ref{e141}), reduces to 
\begin{equation}
\left. r\,\frac{d\psi}{dr}\right|_{a_-} = \left. r\,\frac{d\psi}{dr}\right|_{a_+}.
\end{equation}
Note that a failure to satisfy the pressure balance matching condition is associated with a gradient discontinuity in $\psi(r)$ at the plasma
boundary. Such a discontinuity implies the existence of a perturbed current sheet flowing on the boundary. 

Inside the plasma, Newcomb's equation, (\ref{e146}), can be re-expressed in the form\,\cite{slinky}
\begin{equation}\label{e175}
(\hat{f}\,\psi')' - \hat{g}\,\psi = 0,
\end{equation}
where
\begin{align}
\hat{f}(r) &= \frac{1}{r\,k_0^{\,2}},\\[0.5ex]
\hat{g}(r)&= \frac{1}{r} + \frac{\sigma'\,G}{r\,k_0^{\,2}\,F} -\frac{2\,m\,k\,\sigma}{r^3\,k_0^{\,4}}-\frac{\sigma^2}{r\,k_0^{\,2}}.
\end{align}
Note that Eq.~(\ref{e175}) is singular at any equilibrium magnetic flux-surface, $r=r_s$, lying within the plasma, that satisfies $F(r_s)=0$. An ideal solution (which is unable to reconnect magnetic
field-lines) must satisfy $\psi(r_s)=0$ at such a surface.\cite{freid1,fit}
It is helpful to define
\begin{equation}\label{e178}
c_p = \left(r\,\frac{d\psi_p}{dr}\right)_a,
\end{equation}
where $\psi_p(r)$ is a solution of Eq.~(\ref{e175}) that is well-behaved at $r=0$, satisfies $\psi(r_s)=0$ at any singular surfaces  within the plasma,
and is such that $\psi_p(a)=1$. 
Thus, the solution in the region $0\leq r\leq a$ becomes 
\begin{equation}
\psi(r)= \psi_a\,\psi_p(r),
\end{equation}
where $\psi_a=\psi(a)$, which is the value of the perturbed poloidal magnetic flux at the plasma boundary, is undetermined. 

Outside the plasma, in the region $r>a$, we can write
\begin{equation}
\psi(r)= \psi_a\,\psi_{rwm}(r)= \psi_a\,[c_1\,\psi_{nw}(r)+ c_2\,\psi_{pw}(r)],
\end{equation}
where $\psi_{nw}(a)= \psi_{pw}(a) = 1$, and $c_1+c_2=1$. [See Eqs.~(\ref{e187}) and (\ref{e188}).] This automatically satisfies the matching condition (\ref{e175}). Here, 
\begin{align}
\psi_{nw}(r) &= \frac{r}{a}\,\frac{K_{|m|}'(k\,r)}{K_{|m|}'(k\,a)},\\[0.5ex]
\psi_{pw}(r)&= \frac{r}{a}\,\frac{I_{|m|}'(k\,r)\,K_{|m|}'(k\,b)-K_{|m|}'(k\,r)\,I_{|m|}'(k\,b)}{I_{|m|}'(k\,a)\,K_{|m|}'(k\,b)-K_{|m|}'(k\,a)\,I_{|m|}'(k\,b)}.
\end{align}
Note that $\psi_{nw}(\infty)=\psi_{pw}(b)=0$, in accordance with Eqs.~(\ref{e149}) and (\ref{e153}). Furthermore, it is
understood that $\psi_{pw}(r>b)=0$. 
It is helpful to define
\begin{align}
c_{nw} &= \left(r\,\frac{d\psi_{nw}}{dr}\right)_a= \frac{m^2+(k\,a)^2}{k\,a}\, \frac{K_{|m|}(k\,a)}{K_{|m|}'(k\,a)},\\[0.5ex]
c_{pw}&=  \left(r\,\frac{d\psi_{pw}}{dr}\right)_a=  \frac{m^2+(k\,a)^2}{k\,a}\, \frac{I_{|m|}(k\,a)\,K_{|m|}'(k\,b)-K_{|m|}(k\,a)\,I_{|m|}'(k\,b)}{I_{|m|}'(k\,a)\,K_{|m|}'(k\,b)-K_{|m|}'(k\,a)\,I_{|m|}'(k\,b)}.
\end{align}
It is easily demonstrated that
\begin{align}
\delta W_{nw} &= \frac{\pi^2\,R_0}{\mu_0}\,\frac{\psi_a^{\,2}}{m^2+(k\,a)^2}\left(c_p - c_{nw}\right),\\[0.5ex]
\delta W_{pw} &= \frac{\pi^2\,R_0}{\mu_0}\,\frac{\psi_a^{\,2}}{m^2+(k\,a)^2}\left(c_p -c_{pw}\right).
\end{align}
Furthermore,
\begin{align}\label{e187}
c_1&= \frac{c_p-c_{pw}}{c_{nw}-c_{pw}},\\[0.5ex]
c_2&= \frac{c_{nw}-c_{p}}{c_{nw}-c_{pw}}.\label{e188}
\end{align}
Note that $c_1+c_2=1$. 

The resistive wall dispersion relation becomes
\begin{equation}
\sqrt{\frac{\gamma\,\tau_w}{\delta_w}}\tanh\!\sqrt{\delta_w\,\gamma\,\tau_w} = c_b\left(\frac{c_{nw}-c_p}{c_p-c_{pw}}\right),
\end{equation}
where
\begin{equation}
c_b =\frac{(k\,b)^2}{m^2+(k\,b)^2}\,\,\frac{K_{|m|}'(k\,b)}{K_{|m|}'(k\,a)}\,\left[I_{|m|}'(k\,a)\,K_{|m|}'(k\,b)-K_{|m|}'(k\,a)\,I_{|m|}'(k\,b)\right].
\end{equation}
Thus, the whole problem is fully specified, for given poloidal and toroidal mode numbers,  once the parameter $c_p$, defined in Eq.~(\ref{e178}),  is numerically calculated from the modified Newcomb
equation, (\ref{e175}). 

\subsection{Example Calculation}
\begin{figure}
\centerline{\includegraphics[width=0.8\textwidth]{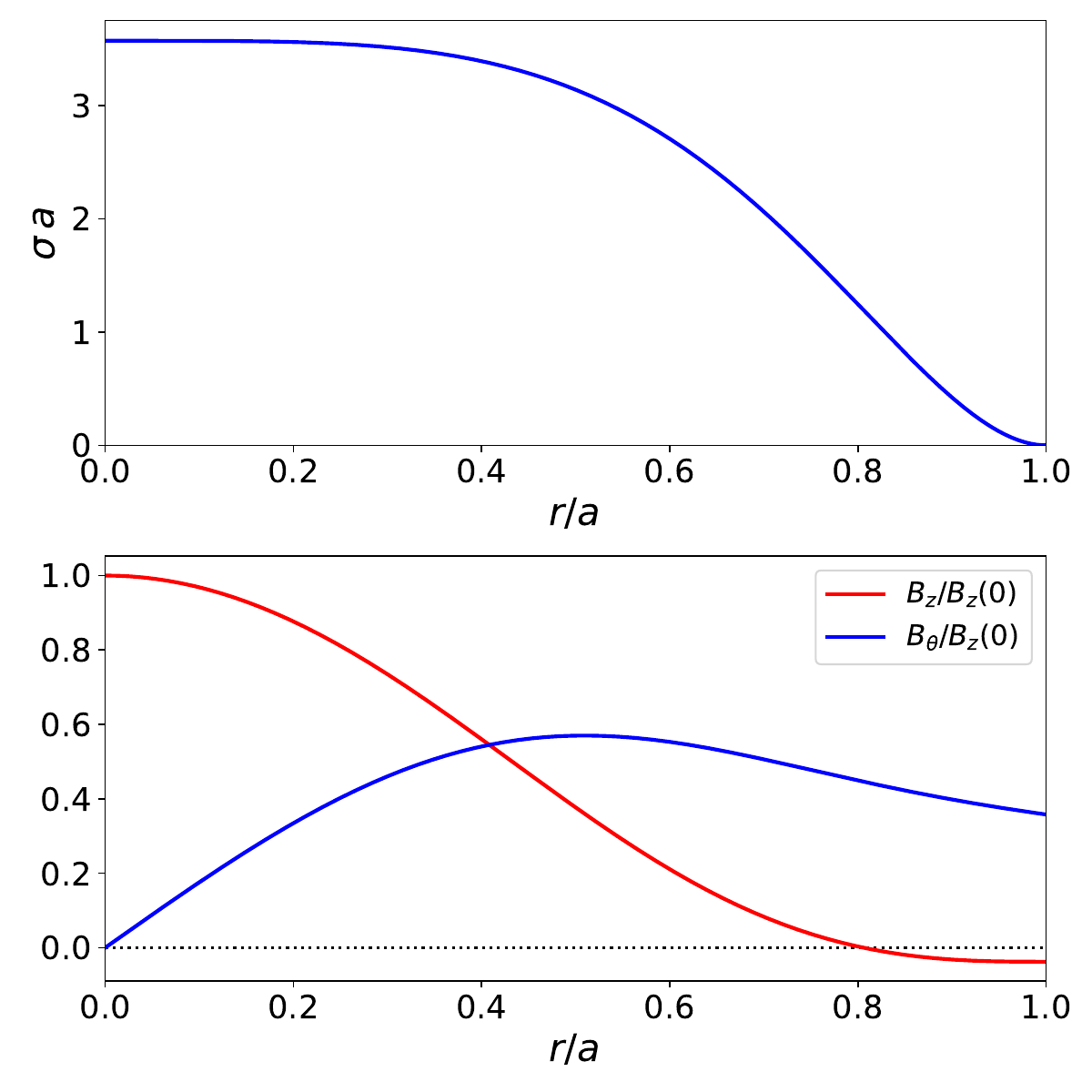}}
\caption{Force-free reversed-field pinch equilibrium characterized by $a/R_0 = 0.25$, $\sigma_0\,a=3.57$, $\alpha= 4.0$, and $\nu=2.0$.}\label{fig3}
\end{figure}

Let us adopt the following equilibrium parameters: $a/R_0 = 0.25$, $\sigma_0\,a=3.57$, $\alpha= 4.0$, and $\nu=2.0$. The resulting  generic reversed-field pinch equilibrium is
shown in Fig.~\ref{fig3}. The characteristic pinch and reversal parameters\,\cite{freid1}  are ${\mit\Theta}= 1.70$ and $F = -0.18$, respectively. As is well-known, it is necessary to place the wall relatively close to a reversed-field pinch plasma in order to stabilize all possible ideal external-kink
modes.\cite{freid1} In the present case, we choose $b/a=1.12$. The thickness of the wall is $d/a=0.4$, which is the largest value that we could adopt
and plausibly argue that $d/b\ll 1$. 

Consider the $m=-1/n=11$ resistive wall mode. This is a mode that possesses a resonant surface in the plasma located at $r/a= 0.463$. The no-wall and perfect-wall energies
of the mode are $\delta W_{nw} = 0.359\,(\pi^2\,R_0\,\psi_a^{\,2}/\mu_0)$ and $\delta W_{pw}= 1.030\,(\pi^2\,R_0\,\psi_a^{\,2}/\mu_0)$, respectively. 
The fact that both energies are positive indicates that the mode is stable. In fact, the decay-rate of the mode is $-\gamma\,\tau_w= 0.02677$. 
If the wall were in the thin-shell limit, but had the same $\tau_w$ value, then the decay-rate would have been $-\gamma\tau_w=0.02685$. Thus,
the finite thickness of the wall has decreased the decay-rate of mode, in accordance with the discussion in Sect.~\ref{growth}. However, despite the
comparatively large wall thickness, the
reduction is extremely modest. 

\begin{figure}
\centerline{\includegraphics[width=\textwidth]{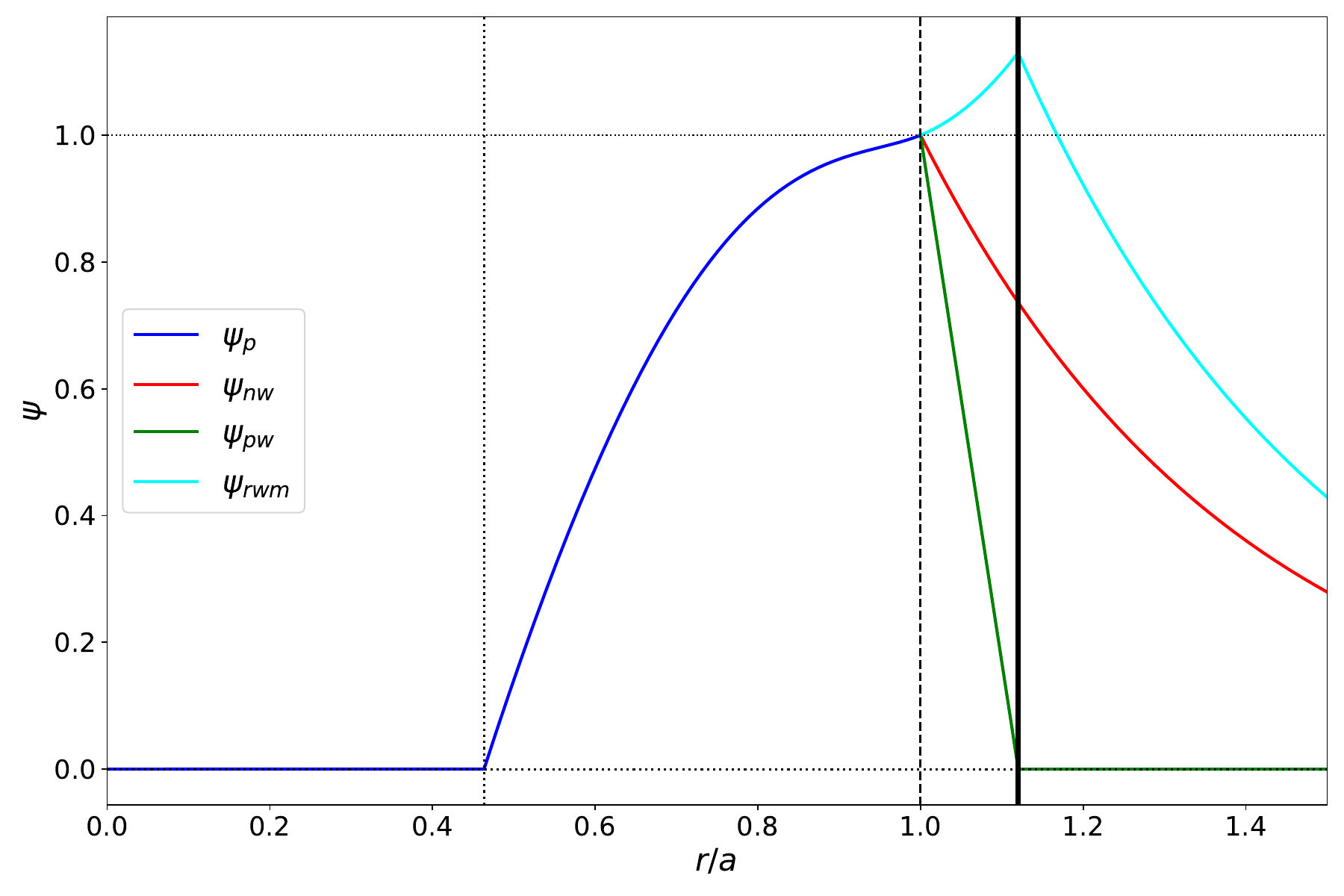}}
\caption{Eigenfunctions of an $m=-1/n=11$ resistive wall mode in a force-free reversed-field pinch equilibrium characterized by $a/R_0 = 0.25$, $\sigma_0\,a=3.57$, $\alpha= 4.0$, $\nu=2.0$, $b/a=1.12$, and $d/a=0.4$. The dotted vertical line indicates the resonant surface, the
dashed vertical line indicates the plasma boundary, and the solid vertical line indicates the inner surface of the wall. }\label{fig4}
\end{figure}

Figure~\ref{fig4} shows the eigenfunctions of the $m=-1/n=11$ resistive wall mode. The no-wall ideal external-kink mode has the
eigenfunction $[\psi_p(r)$, $\psi_{nw}(r)]$, where the first function corresponds to the plasma, whereas the second corresponds to
the vacuum. It can be seen that this eigenfunction has a gradient discontinuity at the plasma boundary, indicating that
it does not satisfy the pressure balance matching condition. 
The perfect-wall ideal external-kink mode has the
eigenfunction $[\psi_p(r)$, $\psi_{pw}(r)]$. Again, this eigenfunction has a gradient discontinuity at the plasma boundary, indicating that
it does not satisfy the pressure balance matching condition. Finally, the resistive wall mode has the eigenfunction $[\psi_p(r)$, $\psi_{rwm}(r)]$. 
Note that this eigenfunction is completely continuous across the plasma boundary, indicating that it  does satisfy the pressure balance matching condition. 

\begin{figure}
\centerline{\includegraphics[width=0.9\textwidth]{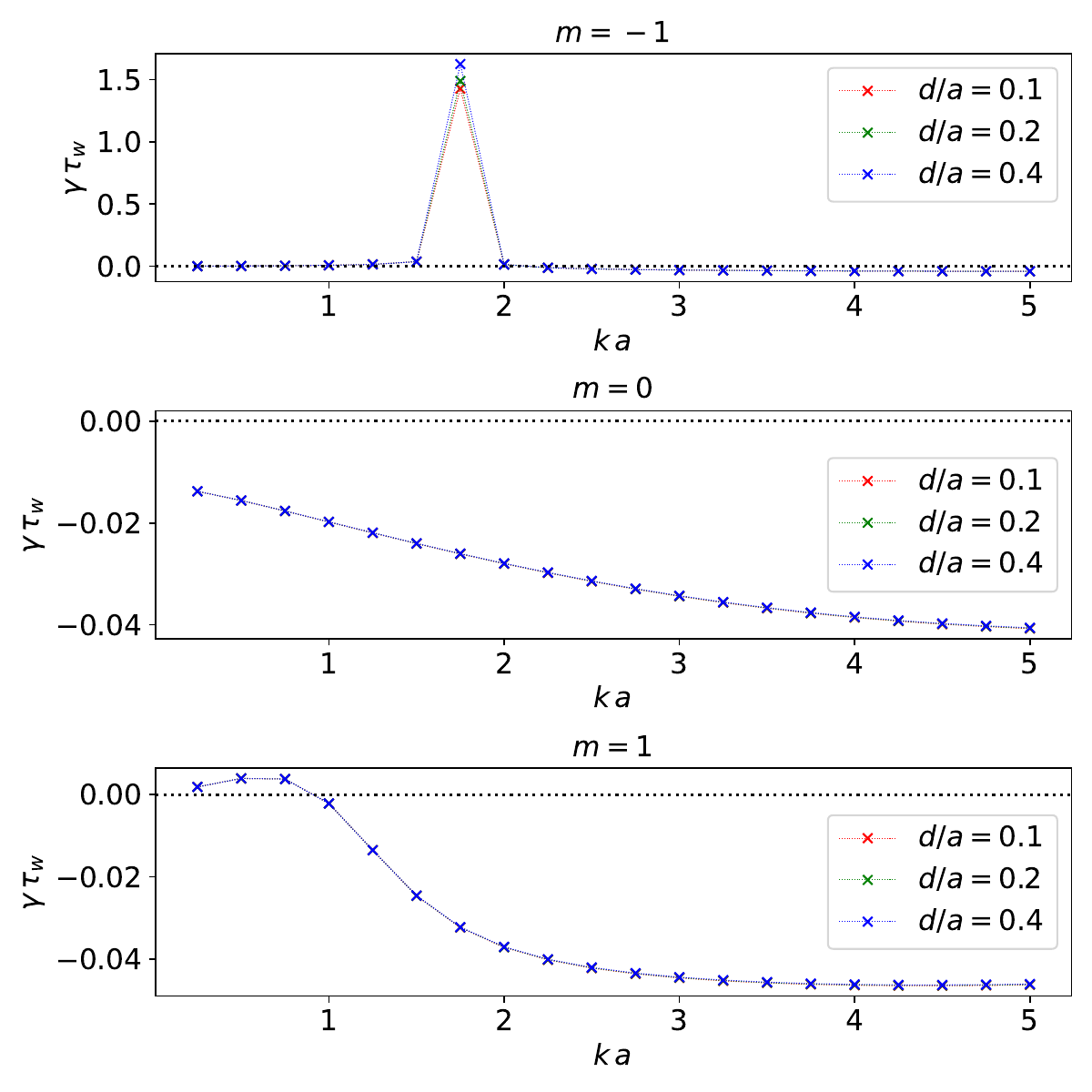}}
\caption{Growth-rates of  $m=-1$, $m=0$, and $m=1$ resistive wall modes calculated in a force-free reversed-field pinch equilibrium characterized by $a/R_0 = 0.25$, $\sigma_0\,a=3.57$, $\alpha= 4.0$, $\nu=2.0$, and $b/a=1.12$, for various values of the wall thickness.  Note that most of the data points
for different wall thicknesses plot on top of one another.}\label{fig5}
\end{figure}

Figure~\ref{fig5} shows the growth-rates of the $m=-1$, $m=0$, and $m=1$ resistive wall modes, calculated for $n$ in the range 1 to 20, and
for various values of the wall thickness. It can be seen that the varying wall thickness makes no discernible difference to the growth-rates of
the modes, with the exception of the $m=-1/n=7$ mode. It turns out that the $m=-1/n=7$ ideal external-kink mode is barely stabilized by a perfectly-conducting wall located at
$b/a=1.12$. 
Consequently, the corresponding resistive wall mode has a comparatively large growth-rate. In fact, in this case, it can be seen that increasing
wall thickness (at fixed $\tau_w$) leads to a noticeable   increase in the growth-rate, in accordance with the discussion in Sect.~\ref{growth}. 
Thus, we conclude that thick-wall effects are only important for resistive wall modes that lie fairly close to the perfect-wall ideal
stability boundary. 

\section{Rotational Stabilization of Resistive Wall Mode}\label{hub}
\subsection{Generalized Hu-Betti Formula}
So far, the analysis presented this paper suggests that the marginal stability point for the
resistive wall mode is the same as that for the no-wall ideal external-kink mode: that is, $\delta W_{pw}=0$. 
In other words, a close-fitting resistive wall is capable of transforming a rapidly growing ideal external-kink mode
into a slowly growing resistive wall mode, but is unable to completely stabilize the mode.\cite{freid1} This conclusion is ultimately a consequence of the
 fact that the plasma potential energy, $\delta W_p$, is a real quantity. However, it turns out that resonances within the plasma, in combination with toroidal plasma rotation,  allow 
$\delta W_p$ to acquire an imaginary component.\cite{res1} The resonances in question include resonances with the sound wave
continuum,\cite{res2} resonances with the shear-Alfv\'{e}n wave continuum,\cite{res3} and resonances with trapped and circulating
particles.\cite{res4,res5}  Furthermore, above a critical plasma rotation rate, the resistive wall mode is stabilized by the imaginary component of $\delta W_p$.\cite{bond}

Let us write
\begin{equation}
\delta W_p = \delta W_p^{(r)} + {\rm i}\,\delta W_p^{(i)},
\end{equation}
where $\delta W_p^{(r)}$ and $\delta W_p^{(i)}$ are the real and imaginary components of $\delta W_p$, respectively. 
We can then make the following redefinitions:
\begin{align}
\delta W_{pw} &=\delta W_p^{(r)} +\delta W_s+\delta W_v^{(b)},\\[0.5ex]
\delta W_{nw} &=\delta W_p^{(r)} +\delta W_s+\delta W_v^{(\infty)}.
\end{align}
Incidentally, it is obvious, from their definitions, that  $\delta W_s$, $\delta W_v^{(b)}$, and $\delta W_v^{(\infty)}$ are all real quantities. 
Note that the real part of the resonant contribution to $\delta W_p$ has been incorporated into $\delta W_p^{(r)}$. Generally
speaking, we expect $\delta W_p^{(i)}$ to be proportional to the toroidal plasma rotation at the resonant point within the plasma.\cite{res2}

In the presence of resonances, our generalized Haney-Freidberg formula, (\ref{hf1}), generalizes further to give 
\begin{equation} \label{e194}
\sqrt{\frac{\gamma\,\bar{\tau}_w}{\bar{\delta}_w}}\,\tanh\left(\!\sqrt{\bar{\delta}_w\,\gamma\,\bar{\tau}_w}\right)= - 
\left(\frac{\delta W_{nw}+ {\rm i}\,\delta W_p^{(i)}}{\delta W_{pw} + {\rm i}\,\delta W_p^{(i)}}\right),
\end{equation}
where $\gamma = \gamma_r -{\rm i}\,\omega_r$. Here, $\gamma_r$ and $\omega_r$ are the real growth-rate and real frequency of the resistive
wall mode, respectively. 
The previous formula is a generalization of a formula that first appeared in a paper by Hu \& Betti in 2004.\cite{res4}

Note, incidentally,  that it is not obvious that the force operator, ${\bf F}(\bxi)$, remains self-adjoint in the presence of  an imaginary resonant contribution to $\delta W_p$. 
Given that the proof of the variation principle,  upon which Eq.~(\ref{hf1})  is based, is itself based on the self-adjoint nature of the force operator, this calls into question the validity of
the Hu \& Betti formula, and the previous generalization. However, in the following, we shall assume that these formulae are correct. 

\subsection{Marginal Stability Point}\label{smarg}
Let us assume that the resistive wall mode is unstable in the absence of an imaginary resonant contribution to $\delta W_p$, which implies that
$\delta W_{nw} <0$ and $\delta W_{pw}>0$. Let us search for a marginal stability point at which $\gamma_r=0$. If
we define
\begin{align}
x&= \omega_r\,\bar{\tau}_w,\\[0.5ex]
y &= \frac{\delta W_p^{(i)}}{\delta W_{pw}},\\[0.5ex]
z &= - \frac{\delta W_{nw}}{\delta W_{pw}},
\end{align}
$\zeta= [x/(2\,\bar{\delta}_w)]^{1/2}$,
$\mu = (2\,\bar{\delta}_w\,x)^{1/2}$,
$S = \sinh\mu$, 
$C=\cosh\mu$,
$s = \sin\mu$, 
$c=\cos\mu$,
and
\begin{align}
\alpha(x)&= \zeta\left(\frac{S-s}{C+c}\right),\\[0.5ex]
\beta(x) &= \zeta\left(\frac{S+s}{C+c}\right),
\end{align}
then Eq.~(\ref{e194})
yields
\begin{align}
\alpha&= \frac{z-y^2}{1+y^2},\\[0.5ex]
\beta&= \frac{y\,(1+z)}{1+y^2}.
\end{align}
Here, we are assuming, without loss of generality, that $x>0$ and $y>0$. This assumption simply implies an arbitrary  choice of the direction of plasma rotation. 
The previous two equations can be combined to give
\begin{equation}
F(x)\equiv \beta^2-(z-\alpha)\,(1+\alpha) = 0.
\end{equation}
Once $x$ has been determined, by finding the root of the previous equation, $y$ is given by
\begin{equation}
y = \sqrt{\frac{z-\alpha}{1+\alpha}}.
\end{equation}

\begin{figure}
\centerline{\includegraphics[width=0.9\textwidth]{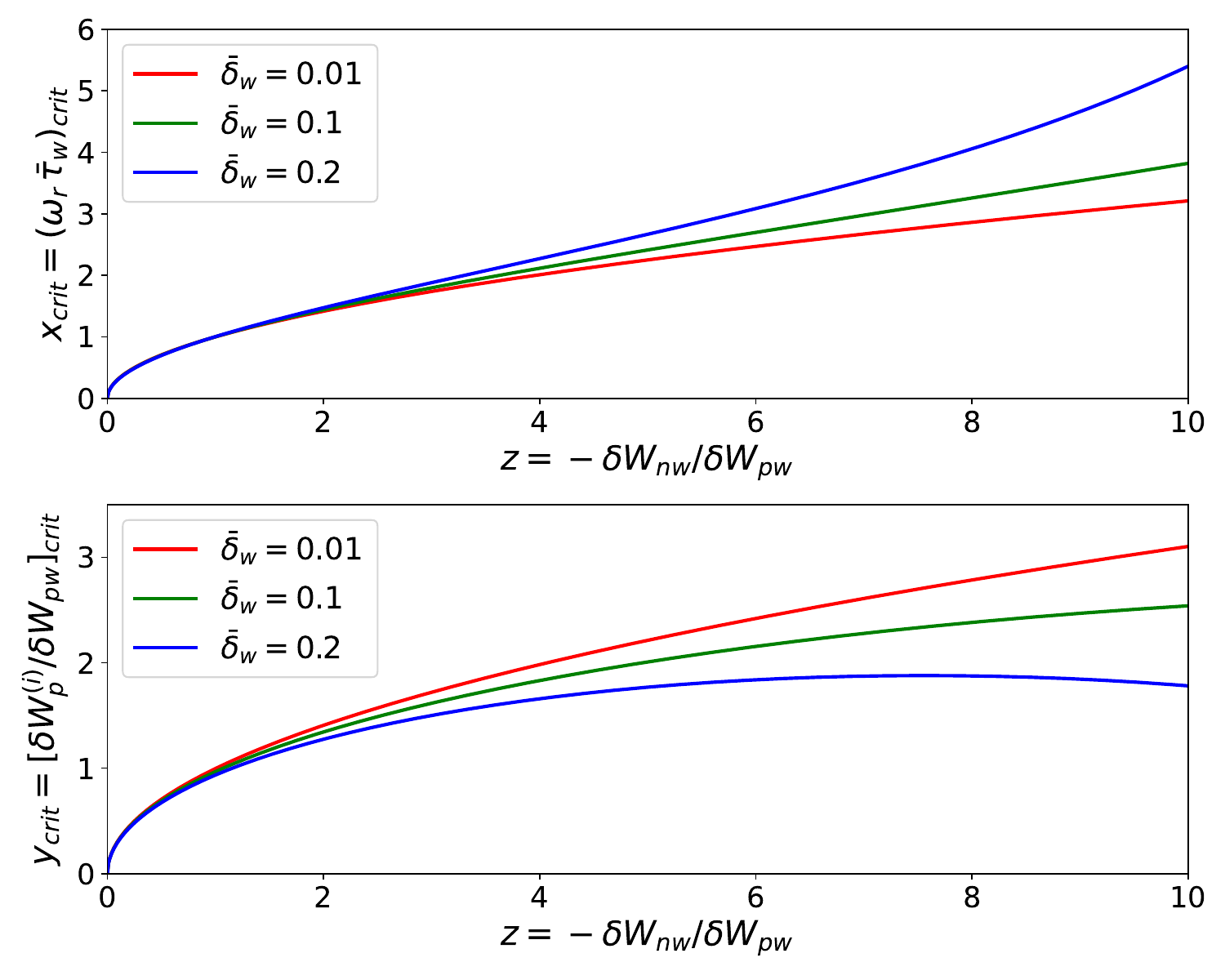}}
\caption{Critical real frequency, $\omega_r$, and critical imaginary part of the plasma potential energy, $\delta W_p^{(i)}$, needed to stabilize the resistive wall
mode according to the generalized Hu-Betti formula, (\ref{e194}).}\label{fig6}
\end{figure}

Figure~\ref{fig6} shows the critical real frequency, $\omega_r$, and the  critical imaginary part of the plasma potential energy, $\delta W_p^{(i)}$, needed to stabilize the resistive wall
mode, according to the generalized Hu-Betti formula, (\ref{e194}). It can be seen that increasing wall thickness (at fixed $\bar{\tau}_w$) facilitates the
stabilization of the resistive wall mode, because it decreases the critical value of $\delta W_p^{(i)}$, which corresponds to a decreased critical
plasma rotational rate. On the other hand, the real frequency of the mode at the marginal stability point increases with increasing wall thickness.
Note, finally, that the thick-wall stabilization criterion only differs substantially from the thin-wall stabilization criterion, which is
\begin{equation}
 (\omega_r\,\bar{\tau}_w)_{crit}=\left[\frac{\delta W_p^{(i)}}{\delta W_{pw}}\right]_{crit} = \sqrt{ - \frac{\delta W_{nw}}{\delta W_{pw}}},
 \end{equation}
 when $- \delta W_{nw}/\delta W_{pw}\gtrsim 1$, which implies that the corresponding external-kink mode lies fairly close to the perfect-wall stability boundary.

\subsection{Toroidal Electromagnetic Torque}
As pointed out by J.-K.~Park,\cite{park} there is an intimate connection between the imaginary component of the plasma potential energy, $\delta W_p^{(i)}$,
and the net toroidal electromagnetic torque exerted by the resistive wall on the plasma. Let us investigate this connection. 

Employing the cylindrical
analysis of Sect.~\ref{scyl}, the net outward flux of toroidal angular momentum across a magnetic flux-surface of minor radius $r$, lying
outside the plasma, is\,\cite{tj}
\begin{equation}
T_\phi(r) = \frac{1}{2\,\mu_0}\oint\int_0^{2\pi\,R_0} r\,R_0\,(\hat{Q}_z^\ast\,\hat{Q}_r+\hat{Q}_z\,\hat{Q}_r^\ast)\,d\theta\,dz.
\end{equation}
Given that $\hat{\bf Q}={\rm i}\,\nabla V$ in a vacuum region, we deduce that
\begin{equation}\label{e205}
T_\phi(r) = \frac{\pi^2\,n\,R_0}{\mu_0}\,{\rm i}\left(-V^\ast\,r\,\frac{dV}{dr}+V\,r\,\frac{dV^\ast}{dr}\right),
\end{equation}
where the additional factor of $1/2$ comes from averaging $\cos^2(m\,\theta+k\,z)$, and $n=k\,R$ is the toroidal mode number. 
However, it is clear from Eq.~(\ref{e147}) that
\begin{equation}
\frac{d}{dr}\!\left(V_\ast\,r\,\frac{dV}{dr} - V\,r\,\frac{dV^\ast}{dr}\right)=0.
\end{equation}
Hence, we conclude that 
\begin{equation}
\frac{dT_\phi}{dr} =0.
\end{equation}
 in any vacuum region.\cite{t3}

Now, in the vacuum region beyond the wall, 
\begin{equation}\label{e208}
V(r)= \psi_a\,c_3\,\hat{V}_{nw}(r)
\end{equation}
[see Eq.~(\ref{e86})] where $\psi_a= (r\,\hat{F}\,\xi)_a$ is the value of the perturbed poloidal magnetic flux at the plasma boundary,
and
\begin{equation}
\hat{V}_{nw}(r) = \frac{1}{k\,a}\,\frac{K_{|m|}(k\,r)}{K_{|m|'}(k\,a)}.
\end{equation}
[See Eq.~(\ref{e150}).] 
However, it is obvious from Eqs.~(\ref{e205}) and (\ref{e208}) that 
\begin{equation}
T_\phi(r>b)= 0.
\end{equation}
 In other words, the net flux of toroidal angular
momentum from the plasma-wall system is zero. 

Now, in the vacuum region between the plasma and the wall, 
\begin{equation}\label{e210}
V(r)= \psi_a\left[c_1\,\hat{V}_{nw}(r)+ c_2\,\hat{V}_{pw}(r)\right]
\end{equation}
[see Eq.~(\ref{e85})], where $c_1+c_2=0$. [See Eq.~(\ref{e89}).]  Here, 
\begin{equation}
\hat{V}_{pw}(r)= \frac{1}{k\,a}\,\frac{I_{|m|}(k\,r)\,K_{|m|}'(k\,b)-K_{|m|}(k\,r)\,I_{|m|}'(k\,b)}{I_{|m|}'(k\,a)\,K_{|m|}'(k\,b)-K_{|m|}'(k\,a)\,I_{|m|}'(k\,b)}.
\end{equation}
[See Eq.~(\ref{e154}).]
Note that 
\begin{equation}
\left(r\,\frac{d\hat{V}_{nw}}{dr}\right)_a= \left(r\,\frac{d\hat{V}_{pw}}{dr}\right)_a = 1.
\end{equation}
It follows from Eqs.~(\ref{e205}) and (\ref{e210}) that
\begin{equation}\label{e213}
T_\phi(a) = -\frac{2\pi^2\,R_0\,n\,\psi_a^{\,2}}{\mu_0} \,(\hat{V}_{nw}-\hat{V}_{pm})_a\,{\rm Im}(c_1).
\end{equation}
Thus, if $c_1$ possesses an imaginary component then there is a constant flux of toroidal electromagnetic angular momentum between the plasma and the wall.
This flux is associated with a toroidal electromagnetic torque exerted on the plasma, and an equal and opposite torque exerted on the wall. 

Now, according to Eqs.~(\ref{esurf}), (\ref{e151}), (\ref{e156}), (\ref{e163}), and (\ref{e164}),
\begin{equation}
c_1 = \frac{\mu_0}{\pi^2\,R_0\,\psi_a^{\,2}}\,\frac{\delta W_p + \delta W_s+\delta W_v^{(\infty)}}{(\hat{V}_{nw}-\hat{V}_{pw})_a},
\end{equation}
which implies that
\begin{equation}
{\rm Im}(c_1) = \frac{\mu_0}{\pi^2\,R_0\,\psi_a^{\,2}}\,\frac{{\rm Im}(\delta W_p)}{(\hat{V}_{nw}-\hat{V}_{pw})_a}= \frac{\mu_0}{\pi^2\,R_0\,\psi_a^{\,2}}\,\frac{\delta W_p^{(i)}}{(\hat{V}_{nw}-\hat{V}_{pw})_a}.
\end{equation}
Here, we have made use of the fact that $\delta W_s$ and $\delta W_v^{(\infty)}$ are obviously real quantities. 
Hence, we deduce from Eq.~(\ref{e213}) that\,\cite{park}
\begin{equation}
T_\phi= -2\,n\,\delta W_p^{(i)}.
\end{equation}
Here, $T_\phi$ is the toroidal electromagnetic torque acting on the plasma. 
It follows that the variable $y$, appearing in the analysis of Sect.~\ref{smarg}, represents a normalized electromagnetic torque exerted by the wall on the plasma.
In fact,
\begin{equation}
y = - \frac{T_{\phi}}{2\,n\,\delta W_{pw}}.
\end{equation}
Thus, we can reinterpret Fig.~\ref{fig6} as stating, firstly,  that the rotational stabilization of the resistive wall mode requires the assistance of such a torque, 
and, secondly, that the critical torque needed to stabilize the mode decreases with increasing wall thickness (at constant $\tau_w$). 

\section{Summary}
This paper investigates the external-kink stability of a toroidal plasma surrounded by a rigid resistive wall. 
The main aim of this paper is to extend the well-known analysis of Haney \& Freidberg\,\cite{hf} to allow for a  wall that is sufficiently
thick that the thin-shell approximation does not necessarily hold. First, in Sect.~\ref{energy}, we demonstrate that the MHD force operator remains self-adjoint
in the presence of a thick resistive wall. Next, in Sect.~\ref{svar}, making use of the self-adjoint property, 
we modify the variational principle of Haney \& Freidberg to allow for a thick wall. Finally, in Sect.~\ref{minimize}, we minimize the
plasma potential energy to obtain a generalized  Haney-Freidberg formula, (\ref{hf1}) and (\ref{hf2}), for the growth-rate of the resistive wall mode that allows the wall to lie
either in the thin-shell regime, the thick-shell regime, or somewhere in between. We find that thick-wall effects do not
change the marginal stability point of the  mode, but introduce an interesting asymmetry between growing and decaying modes. Growing modes have
growth-rates that exceed those predicted by the original Haney-Freidberg formula. (Here, we are comparing walls with differing thicknesses but the same L/R time.) On the other hand, decaying modes have decay-rates that are less than those predicted by the original 
formula. We can even generalize the Haney-Freidberg formula to allow for walls with varying thickness and electrical conductivity. [See Eq.~(\ref{general}).] 

We also show, during the course of our investigation, that the eigenfunctions conventionally used to calculate the no-wall and the perfect-wall plasma potential energies of
ideal external-kink mode that feature in 
 the Haney-Freidberg formula do not satisfy the pressure balance matching condition at the plasma boundary. We   then explain why this is not
problematic. In particular,  the resistive wall mode eigenfunction is found to satisfy the pressure balance matching condition. 

In Sect.~\ref{scyl}, we perform a cylindrical calculation for a generic force-free reversed-field pinch plasma equilibrium that reveals that thick-wall effects have no
noticeable effect on the growth-rates of the various resistive wall modes to which the plasma is subject, except when the mode is question lies quite close to the perfect-wall stability boundary.
For such a comparatively rapidly growing mode, thick-wall effects perceptibly increase the growth-rate. 

Finally, in Sect.~\ref{hub}, we generalize the  well-known Hu-Betti formula\,\cite{res4} for the rotational stabilization of the resistive wall mode
to take thick-wall effects into account. We find that increasing wall thickness (at fixed L/R time) facilitates the
rotational stabilization of the resistive wall mode, because it decreases the critical toroidal electromagnetic  torque that the
wall must exert on the plasma. On the other hand, the real frequency of the mode at the marginal stability point increases with increasing wall thickness.

\section*{Acknowledgements}
This research was directly funded by the U.S.\ Department of Energy, Office of Science, Office of Fusion Energy Sciences, under  contract DE-SC0021156. 

\section*{Data Availability Statement}
The digital data used in the figures in this paper can be obtained from the author upon reasonable request.

\newpage

\section*{References}
\begin{thebibliography}{99}\baselineskip 5ex

\bibitem{ek1} G.~Laval, R.~Pellat and J.S.~Soule, Phys.\ Fluids {\bf 17}, 815 (1974). 

\bibitem{ek2} R.L.~Dewar, R.C.~Grimm, J.L.~Johnson, E.A.~Frieman, J.M.~Greene and P.H.~Rutherford, Phys.\ Fluids {\bf 17}, 930 (1974).

\bibitem{ek3} F.A.~Haas, Nucl.\ Fusion {\bf 15}, 407 (1975).

\bibitem{tok} J.A.~Wesson, {\em Tokamaks}, 4th Ed. (Oxford University Press, Oxford UK, 2011).

\bibitem{fit} R.~Fitzpatrick, {\em Tearing Mode Dynamics in Tokamak Plasmas}. (IOP, Bristol UK, 2023).

\bibitem{rwm1} D.~Pfirsch and H.~Tasso, Nucl.\ Fusion {\bf 11}, 259 (1971).

\bibitem{rwm2} J.P.~Goedbloed, D.~Pfirsch and H.~Tasso, Nucl.\ Fusion {\bf 12}, 649 (1972). 

\bibitem{bern} I.B.~Bernstein, E.A.~Frieman, M.D.~Kruskal  and R.M.~Kulsrud, Proc.\ R.\ Soc.\ London, Ser.\ A {\bf 244}, 17 (1958).

\bibitem{freid} J.P.~Freidberg, Rev.\ Mod.\ Phys.\ {\bf 54}, 801 (1982).

\bibitem{freid1} J.P.~Freidberg,  {\em Ideal Magnetohydrodynamics}. (Plenum, New York NY, 1987).

\bibitem{goed} J.P.~Goedbloed and S.~Poedts, {\em Principles of Magnetohydrodynamics}. (Cambridge University Press, Cambridge UK, 2004).

\bibitem{hf} S.W.~Haney and J.P.~Freidberg, Phys.\ Fluids B {\bf 1}, 1637 (1989). 

\bibitem{thick0} C.G.~Gimblett, Nucl.\ Fusion {\bf 26}, 617 (1986).

\bibitem{thick01} R.~Fitzpatrick, S.C.~Guo, D.J.~Den Hartog and C.C.~Hegna, Phys.\ Plasmas {\bf 6}, 3878 (1999).

\bibitem{chap} B.E.~Chapman, R.~Fitzpatrick, D.~Craig,   P.~Martin and G.~Spizzo, Phys.\ Plasmas {\bf 11}, 2156 (2004).

\bibitem{thick1} L.-J.~Zheng and M.T.~Kotschenreuther, Phys.\ Plasmas {\bf 12}, 072504 (2005).

\bibitem{thick2} V.D.~Pustovitov, Phys.\ Plasmas {\bf 19}, 062503 (2012).

\bibitem{thick3} R.~Fitzpatrick, Phys.\ Plasmas {\bf 20}, 012504 (2013).

\bibitem{thick4} N.D.~Lepikhin and V.D.~Pustovitov, Phys.\ Plasmas {\bf 21}, 042504 (2014).

\bibitem{rfp} R.~Fitzpatrick,  Phys.\ Plasmas {\bf 31}, 042510 (2024).

\bibitem{jackson} J.D.~Jackson,  {\em Classical Electrodynamics}, 3rd Ed. (Wiley \& Sons, Hoboken NJ, 1998).

\bibitem{new} W.A.~Newcomb, Ann.\ Phys.\ (NY) {\bf 10}, 232 (1960).

\bibitem{slinky} R.~Fitzpatrick, Phys.\ Plasmas {\bf 6}, 1168 (1999).

\bibitem{res1} J.W.~Berkery, R.~Betti, Y.Q.~Liu and S.A.~Sabbagh, Phys.\ Plasmas {\bf 30}, 120901 (2023).

\bibitem{res2} R.~Betti and J.P.~Freidberg, Phys.\ Rev.\ Lett.\ {\bf 74}, 2949 (1995).

\bibitem{res3} L.J.~Zheng, M.~Kotschenreuther and M.S.~Chu, Phys.\ Rev.\ Lett.\ {\bf 95}, 255003 (2005).

\bibitem{res4} B.~Hu and R.~Betti, Phys.\ Rev.\ Lett.\ {\bf 93}, 105002 (2004).

\bibitem{res5} B.~Hu, R.~Betti and J.~Manickam,  Phys.\ Plasmas {\bf 12}, 057301 (2005).

\bibitem{bond} A.~Bondeson and D.~Ward, Phys.\ Rev.\ Lett.\ {\bf 72}, 2709 (1994).

\bibitem{park} J.-K.~Park, Phys.\ Plasmas {\bf 18}, 110702 (2011).

\bibitem{tj} R.~Fitzpatrick, {\em Calculation of Tearing Mode Stability in an Inverse Aspect-Ratio
Expanded Tokamak Plasma Equilibrium}, submitted to Physics of Plasmas (2024). 

\bibitem{t3} R.~Fitzpatrick, R.J.~Hastie, T.J.~Martin and C.M.~Roach, Nucl.\ Fusion {\bf 33}, 1533 (1993).

\end{thebibliography}
\end{document}